\documentclass[11pt,letterpaper]{article}
\pdfoutput=1
\usepackage{jheppub}
\usepackage{color}
\usepackage[table]{xcolor}
\usepackage{graphicx}

\usepackage{verbatim}
\usepackage{amsmath}
\usepackage{amssymb}
\usepackage{subfig}
\usepackage{url}
\usepackage{slashed}
\usepackage{xspace}
\usepackage[separate-uncertainty=true]{siunitx}

\usepackage{multirow}
\usepackage{threeparttable}
\usepackage{paralist}

\DeclareRobustCommand{\Fig}[1]{Fig.~\ref{#1}}

\newcommand{\be}{\begin{equation}}
\newcommand{\ee}{\end{equation}}

\newcommand{\pt}{\ensuremath{p_T}\xspace}

\begin{document}
\title{\boldmath Prospects for a measurement of the $W$ boson mass in the all-jets final state at hadron colliders}

\author[a]{Marat Freytsis,}
\author[b]{Philip Harris,}
\author[c]{Andreas Hinzmann,}
\author[d,e]{Ian Moult,}
\author[f]{Nhan Tran}
\author[f]{and Caterina Vernieri}

\affiliation[a]{Institute of Theoretical Science, University of Oregon, Eugene, OR 97403, USA}
\affiliation[b]{Massachusetts Institute of Technology, 77 Massachusetts Ave, Cambridge, MA 02139, USA}
\affiliation[c]{University of Hamburg, Hamburg, Germany}
\affiliation[d]{Berkeley Center for Theoretical Physics, University of California, Berkeley, CA 94720, USA}
\affiliation[e]{Theoretical Physics Group, Lawrence Berkeley National Laboratory, Berkeley, CA 94720, USA}
\affiliation[f]{Fermi National Accelerator Laboratory, Batavia, IL 60510, USA}

\emailAdd{andreas.hinzmann@cern.ch}

\abstract{
Precise measurements of the mass of the $W$ boson are important to test the overall consistency of the Standard Model of particle physics. The current best measurements of the $W$ boson mass come from single production measurements at hadron colliders in its decay mode to a lepton (electron or muon) and a neutrino and pair production of $W$ bosons at lepton colliders, where both the leptonic and hadronic decay modes of the $W$ boson have been considered. In this study, prospects for a measurement of the $W$ boson mass in the all-jet final state at hadron colliders are presented. The feasibility of this measurement takes advantage of numerous recent developments in the field of jet substructure. Compared to other methods for measuring the $W$ mass, a measurement in the all-jets final state would be complementary in methodology and have systematic uncertainties orthogonal to previous measurements. We have estimated the main experimental and theoretical uncertainties affecting a measurement in the all-jet final state. With new trigger strategies, a statistical uncertainty for the measurement of the mass difference between the $Z$ and $W$ bosons of 30 MeV could be reached with HL-LHC data corresponding to 3000 fb$^{-1}$ of integrated luminosity. However, in order to reach that precision, the current understanding of non-perturbative contributions to the invariant mass of $W\to q\bar{q}'$ and $Z\to b\bar{b}$ jets will need to be refined. Similar strategies will also allow the reach for generic boosted resonances searches in hadronic channels to be extended.
}

\begin{flushright}
  FERMILAB-PUB-18-315-E 
\end{flushright}

\maketitle

\section{Introduction}
\label{sec:introduction}

In the Standard Model (SM) of particle of physics, electroweak interactions are mediated by the photon and the $W$ and $Z$ bosons~\cite{Glashow:1961tr,Salam:1964ry,Weinberg:1967tq}. While at lowest order in electroweak theory, the mass of the $W$ boson can be expressed solely as a function of the $Z$ boson mass, the fine-structure constant and the Fermi constant, this statement is modified by higher order corrections, most prominently from other heavy particles in the SM~\cite{Awramik:2003rn,Sirlin:1980nh} and potentially also from particles beyond the SM. Global fits to the SM parameters~\cite{Baak:2014ora}, constraining physics beyond the SM, are currently limited by the precision of the $W$ boson mass measurement. Precise measurements of the mass of the $W$ boson are therefore important to test the overall consistency of the SM. The current best measurements of the $W$ boson mass come from single production measurements at hadron colliders~\cite{Aaboud:2017svj,Aaltonen:2013iut,Aaltonen:2012bp,Abazov:2012bv,Alitti:1991dk,Arnison:1985ut} in its decay mode to a lepton ($e$ or $\mu$) and a neutrino with a branching ratio of \SI{21.34(31)}{\percent}~\cite{Patrignani:2016xqp} and pair production of $W$ bosons at lepton colliders~\cite{Abdallah:2008ad,Schael:2006mz,Achard:2005qy,Abbiendi:2005eq}, where both the leptonic decay mode and the decay mode of the $W$ boson to a $q\bar{q}'$ pair with a branching ratio of \SI{67.41(27)}{\percent} has been considered.
The current world average (not yet considering LHC measurements) is \SI{80.385(15)}{\GeV}~\cite{Patrignani:2016xqp}.

In this paper, we explore the feasibility of a new channel, namely single production at a hadron collider in the decay mode with the highest branching ratio to a $q\bar{q}'$ pair. At hadron colliders, the $W$ boson mass cannot be fully reconstructed in the lepton plus neutrino decay mode. Only its transverse mass can be extracted by estimating the neutrino transverse momentum from the measured missing transverse momentum in the event. The $q\bar{q}'$ decay mode allows for the reconstruction of the full 4-momentum of the $W$ boson through exclusively visible particles. The hadronic decay mode, compared to the lepton plus neutrino decay mode has the potential to avoid the experimental systematic uncertainties related to the measurement of the missing transverse momentum and theoretical uncertainties related to the transverse mass~\cite{CarloniCalame:2016ouw,Bozzi:2015hha}. 
Additionally, the absence of missing transverse energy yields a more narrow peak that is predominantly invariant across a broad kinematic regime of $W$ boson transverse momentum.
Single production of $W\to q\bar{q}'$ results in a rather clean final state, compared to e.g., $t\bar{t}$ production, because the quarks originating from the $W$ boson can form jets of hadrons without color reconnection to other quarks in the event not originating from the $W$ boson.

\begin{figure}[htb]
\begin{center}
\includegraphics[width=0.4\linewidth]{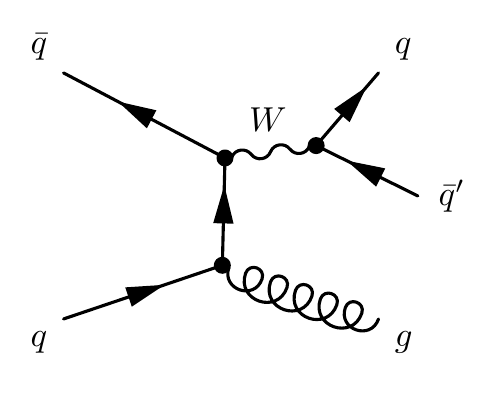}
\end{center}
\caption{Illustrative Feynman diagram showing the production of a $W$ boson together with an additional parton, where the $W$ boson decays to a quark anti-quark pair, $q\bar{q}'$.}
\label{fig:diagram}
\end{figure}

The dominant background to the production of $W\to q\bar{q}'$ at hadron colliders is quantum chromodynamics (QCD) multijet production. This background can be significantly suppressed by requiring jets with high transverse momenta \pt in the event. We therefore propose a measurement of the $W\to q\bar{q}'$ mass produced in association with a high momentum jet as depicted in Fig.~\ref{fig:diagram}. For such high momentum $W$ bosons, the shower of hadrons originating from the quark anti-quark pair merges into a single large radius jet of particles. Multiple techniques have been proposed to analyze the jet substructure of such jets~\cite{Butterworth:2008iy,Thaler:2010tr,Larkoski:2013eya,Dasgupta:2013ihk,Larkoski:2014gra,Larkoski:2014wba,Larkoski:2015kga,Moult:2016cvt} in order to distinguish jets from $W\to q\bar{q}'$ and jets from multijet background. For a review of recent theory and experimental progress in jet substructure, see \cite{Larkoski:2017jix,Asquith:2018igt}. Additionally, such techniques have been shown to reduce theoretical and experimental uncertainties related to the reconstruction of the $W$ boson mass by removing contributions from non-perturbative effects and additional $pp$ interactions happening in the same bunch crossing, so-called pileup interactions. These techniques have been extensively validated by the ATLAS and CMS experiments~\cite{ATLAS-CONF-2015-035, Aad:2015rpa, CMS-PAS-JME-16-003, Khachatryan:2014vla} and it was successfully demonstrated that they can be used to extract the $W\to q\bar{q}'$ mass peak on top of the multijet background~\cite{Sirunyan:2017nvi,Sirunyan:2017dnz,Aad:2015owa}.

Very similar strategies have been used by both ATLAS~\cite{Aaboud:2018zba} and CMS~\cite{Sirunyan:2017dnz,Sirunyan:2017nvi} to place the leading bounds on hadronically decaying resonances beyond the SM in most of the 50--300 GeV range. The dominant systematic uncertainties in these searches result from the selection efficiency and misidentification probability calibrations of the substructure variables used to reduce the multijet backgrounds. These uncertainties are closely correlated with equivalent issues for the $W$ and $Z$ bosons themselves. 
Further, measurements of the Higgs boson decaying to b quarks at high transverse momentum~\cite{Sirunyan:2017dgc} are sensitive to similar systematic effects.
By providing several strategies in which the uncertainties can be studied and quantified in the more well-understood SM channels, improvements can also propagate to analogous measurements and beyond the SM searches, independent of the ultimate precision on the $W$ boson mass that can be achieved. We therefore believe that the extraction of the $W$ boson mass in the all hadronic final state represents a concrete goal that will benefit the field of jet substructure more broadly.

In this paper, we quantify the potential of a measurement of the $W$ boson mass in this new channel at the LHC and the HL-LHC~\cite{Apollinari:2284929} that are expected to deliver proton-proton collision data corresponding to integrated luminosities of 300 fb$^{-1}$ and 3000 fb$^{-1}$, respectively. In Sec.~\ref{sec:simulation} we describe our simulated samples, and simplified detector simulation. In Sec.~\ref{sec:measurement} we present the expected statistical uncertainties on the $W$ boson mass at LHC and HL-LHC as well as trigger strategies. The leading experimental and theoretical uncertainties are discussed in Sec.~\ref{sec:uncertainties}. Finally, we conclude on the feasibility of such a measurement in Sec.~\ref{sec:summary}.

\section{Simulation Setup}
\label{sec:simulation}

Monte Carlo (MC) samples of $W + \text{jets}$ and $Z + \text{jets}$ events, where the $W$ and $Z$ decay into quark anti-quark pairs, as well as multijet events are simulated at a proton-proton center of mass energy of 13 TeV with the leading-order (LO) mode of \textsc{MadGraph5\_aMC@NLO} v5.2.2.2~\cite{Alwall:2014hca,Alwall:2007fs} combined with \textsc{Pythia} version 8.212~\cite{Sjostrand:2014zea} for parton showering with the Monash 2013 tune~\cite{Skands:2014pea}. Additionally, the NNPDF~3.0~\cite{Ball:2014uwa} parton distribution functions (PDF) is used. For cross checks, we use  $W + \text{jets}$ and $Z + \text{jets}$ events produced with \textsc{MadGraph5\_aMC@NLO} combined with \textsc{Herwig++} v2.7.1~\cite{Richardson:2013nfo,Bellm:2013hwb} and its default tune.  Precise predictions of the $W$ and $Z$ boson \pt spectra from Ref.~\cite{Lindert:2017olm} are include in our simulation.
Cross sections are computed at a center of mass energy of 13 TeV. In future runs of LHC and HL-LHC energies up to 14 TeV are foreseen. Conclusions drawn based on the 13 TeV simulation will hold also at 14 TeV as the expected cross section changes from 13 to 14 TeV are only at the 10\%-level for signals and backgrounds.

We employ a detector simulation that reproduces the main resolution effects relevant for jet substructure reconstruction, representative of current and future detector concepts; this simulations employs particle-flow-based reconstruction, such as the CMS~\cite{Sirunyan:2017ulk} or ATLAS~\cite{Aaboud:2017aca} detectors at the LHC.

Due to isospin considerations, jets on average consist of 60\% charged hadrons, 30\% photons (including $\pi_0\to\gamma\gamma$) and 10\% neutral hadrons, although these fractions are subject to large jet-by-jet fluctuations~\cite{Sirunyan:2017ulk,Aaboud:2017aca}. In the simulation, we first categorize the generated particles into charged particles (tracks), photons and neutral hadrons. Tracking inefficiencies occur at high particle momenta and within high momentum jets, where the tracking detector granularity is not sufficient to reconstruct highly collimated particles. Since both inefficiencies are correlated within high momentum jets, they are simulated together by treating charged particles with momenta above a threshold $p_{T,\text{track}}^{\text{max}}=220$ GeV as neutral hadrons.
For jet \pt of 100-500 GeV, the tracking (2-5\%) and HCAL resolutions (5-10\%) are of similar order at CMS. In this range, a generic particle flow algorithm may promote the HCAL measurement over the tracker one.
The threshold $p_{T,\text{track}}^{\text{max}}$ is chosen such that it matches the jet mass resolution of the current CMS detector~\cite{CMS-PAS-JME-14-002} at high momenta, and increased by a factor 2 for the HL-LHC Phase-II upgrade of the CMS tracker~\cite{CMS-TDR-17-001}. The improvement comes from a higher granularity tracking detector which will better distinguish hits from nearby high \pt tracks.
The generated neutral hadrons are then discretized to simulate the spatial resolution of the electromagnetic ($\sigma_{\text{ECAL}}^{\eta}=\sigma_{\text{ECAL}}^{\phi}=0.0175$) and hadronic calorimeters ($\sigma_{\text{HCAL}}^{\eta}=\sigma_{\text{HCAL}}^{\phi}=0.022$). Finally, all particles are smeared according to parametrized resolutions $\sigma_{\text{particle}}^{E}$ for each particle type ($\sigma_{\text{charged particles}}^{\pt}=0.00025 \ \pt/\text{GeV} \oplus 0.015$, $\sigma_{\text{photons}}^{E}=0.021/\sqrt{E/\text{GeV}} \oplus 0.094/(E/\text{GeV}) \oplus 0.005$, $\sigma_{\text{neutral hadrons}}^{E}=0.45/\sqrt{E/\text{GeV}} \oplus 0.05$). The resolutions and granularities have been chosen to match the performance of the CMS detector~\cite{CMS}. The resolution in jet mass and substructure variables for $W\to q\bar{q}'$ and single parton jets with $p_T$ in the range 300 GeV and 3.5 TeV has been compared to CMS public results and found to be compatible~\cite{CMS-PAS-JME-16-003, Khachatryan:2014vla, Sirunyan:2016cao}. The average $W\to q\bar{q}'$ and single parton jet selection efficiencies match with those of CMS~\cite{CMS-PAS-JME-16-003, Khachatryan:2014vla} which are similar to those of ATLAS~\cite{ATLAS-CONF-2015-035, Aad:2015rpa}. As such, while we will present results using simulation generated with the CMS detector configuration, the study is representative for the current and future performance of both experiments.

\section{Measurement Strategy}
\label{sec:measurement}

After introducing the observables used for the measurement, in this section we present two separate strategies to measure the $W$ mass. The first approach is to measure \emph{only} the $W$ boson jet mass peak position $m_W$. The ultimate uncertainty of this approach is constrained experimentally by the jet constituent energy scale calibration. Currently,  it would require a significantly better jet constituent calibration than that achieved by the LHC experiments. The second and more feasible approach is to measure the mass peak position \emph{difference} between the $Z$ boson and the $W$ boson $\Delta m = m_Z-m_W$ because many systematic uncertainties can cancel using the $Z$ boson mass as a standard candle.  Finally, a measurement relying on a recently developed trigger strategy is proposed.

\subsection{Observables}

Jets are clustered from the detector-simulated particles using the anti-$k_T$~\cite{Cacciari:2008gp} algorithm, with a distance parameter of $R = 0.8$. Before computing the invariant mass of the jet, soft radiation is removed iteratively with the modified mass-drop algorithm (mMDT)~\cite{Dasgupta:2013ihk,Butterworth:2008iy}, also known as the soft drop algorithm~\cite{Larkoski:2014wba} with $\beta=0$. The mMDT procedure reduces the mass of quark and gluon jets and improves the mass resolution of $W$ and $Z$ boson jets. Soft drop with the angular exponent $\beta = 0$, soft cutoff threshold $z_\mathrm{cut} = 0.1$, and characteristic radius $R_0 = 0.8$ is applied using the \textsc{FastJet} software package~\cite{Cacciari:2011ma}.
In addition to the soft drop algorithm, we have considered a set of alternative grooming algorithms, namely recursive softdrop~\cite{Dreyer:2018tjj} with the angular exponent $\beta = 1$, soft cutoff threshold $z_\mathrm{cut} = 0.1$, characteristic radius $R_0 = 0.8$, and the number of iterations N set to infinity, trimming~\cite{Krohn:2009th} with subjet size of $R_{\text{sub}}=0.2$ and $f_{\text{cut}=0.03}$ and pruning~\cite{Ellis:2009me} with the soft threshold parameter $z_{\text{cut}}=0.1$ and angular separation threshold of $\Delta R > m_{\text{jet}}/p_{\text{T,jet}}$.

\begin{figure}[tb]
  \begin{center}
    \includegraphics[width=0.65\linewidth]{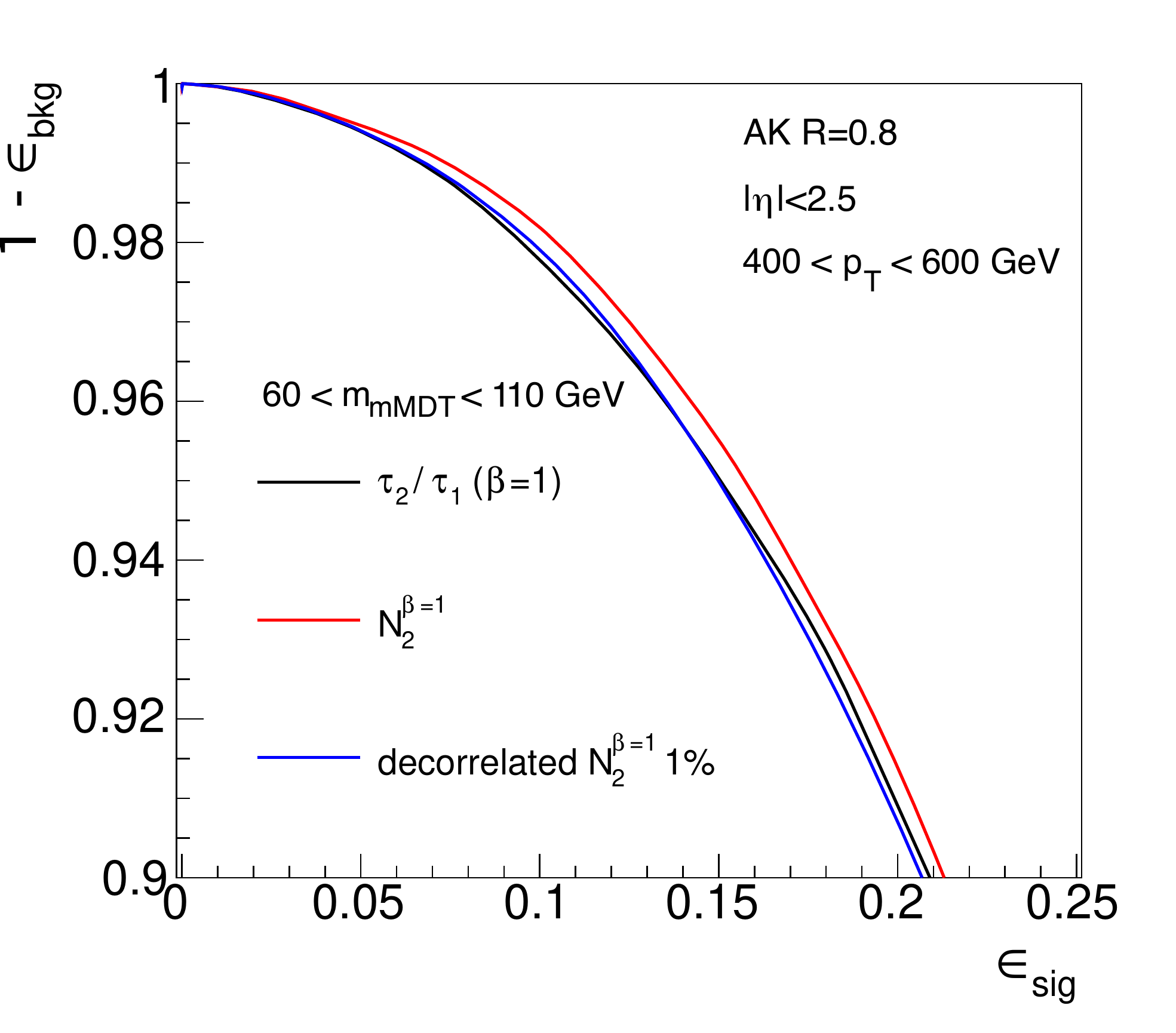}
  \end{center}
  \caption{Quark/gluon jet background rejection power $1-\in_{\text{bkg}}$ as a function of the $W$ jet selection efficiency $\in_{\text{sig}}$ for various substructure observables under study combined with a selection of $60 < m_{\text{mMDT}} < 110$~GeV.}
\label{fig:roc}
\end{figure}

An additional discriminator relying on substructure information is used to further suppress multijet background. Among the most discriminant, $N$-subjettiness observables~\cite{Thaler:2010tr}, $\tau_N$, and the energy correlation function ratios $N_i^\beta$~\cite{Moult:2016cvt} are considered. Here $\tau_{i}$ will take on low values if a jet has $N \ge i$ subjets. The ratio $\tau_2/\tau_1$ therefore discriminates $W \to q\bar{q}'$ jets that contain the shower of a quark anti-quark pair from single quark or gluon jets. Similarly, $N_i^\beta$ attempts to identify $N$-prong jet substructure using information about the energies and pair-wise angles of particles within a jet without requiring a subjet finding procedure. $N_i^\beta$ are ratios of multi-point correlation functions where $\beta$ is again an angular exponent of the pairwise particle distances. $N_2^\beta$ with $\beta=1$ is an observable that distinguish best between quark and gluon jets and intrinsically two-prong jets from a $W$/$Z$ bosons. Its performance in terms of quark and gluon jet background rejection power vs.\ $W$ jet selection efficiency is summarized in Fig.~\ref{fig:roc}, showing similar performance for all considered observables.

\begin{figure}[tb]
  \begin{center}
    \includegraphics[width=0.65\linewidth]{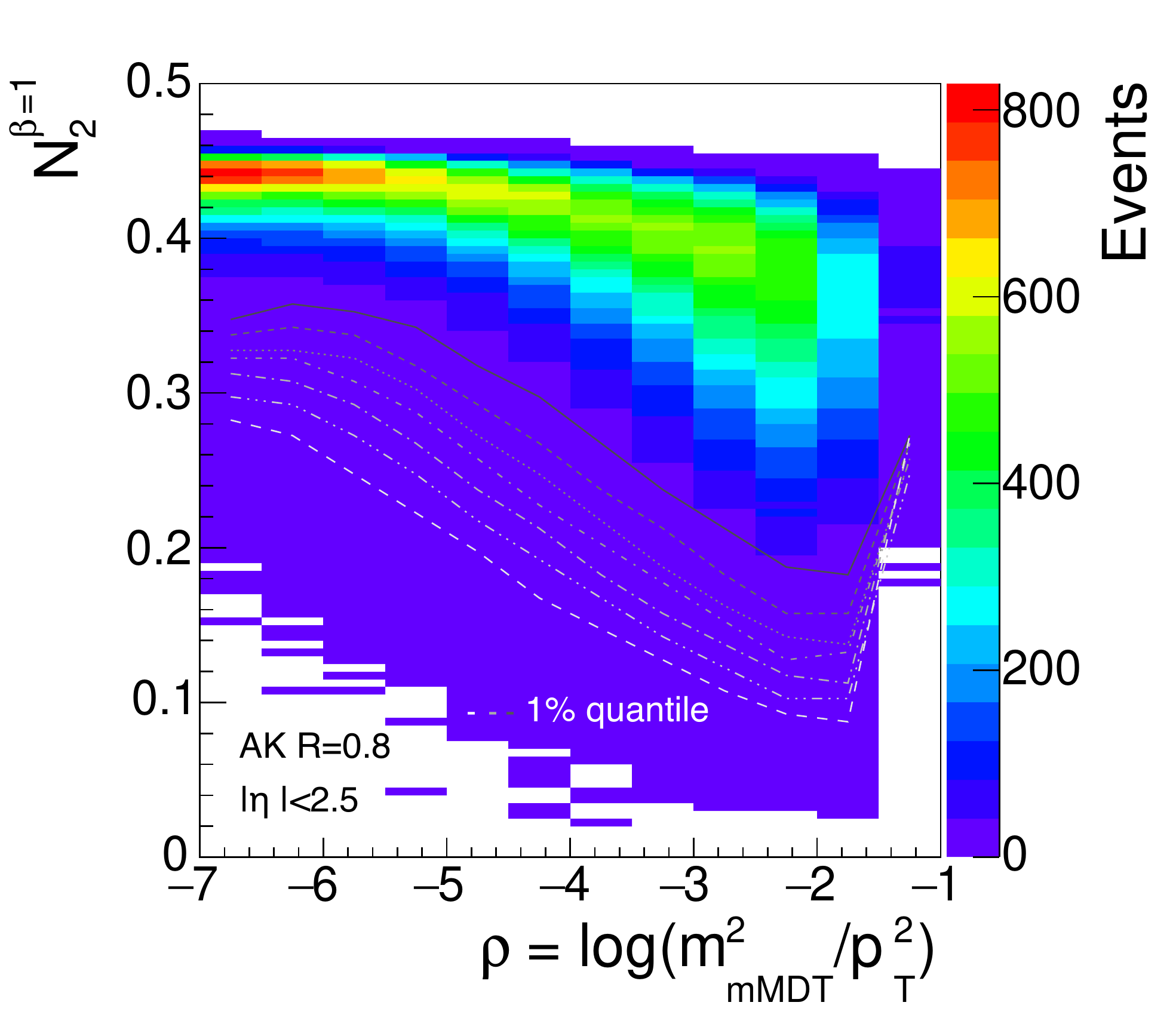}
  \end{center}
  \caption{Number of multijet events as a function $\rho=\log(m_{\text{mMDT}}^2/p_T^2)$ and $N_2^{\beta=1}$ of the leading jet in the event. The lines correspond to the 1\% quantile of the $N_2^{\beta=1}$ distribution in each $\rho$ bin for different jet $p_T$ bins. From top to bottom, the lines correspond to $p_T$ bins with lower borders at 200, 300, 400, 500, 600, 700 and 1000 GeV.}
\label{fig:ddt}
\end{figure}

\begin{figure}[htb]
\begin{center}
\includegraphics[width=0.45\linewidth]{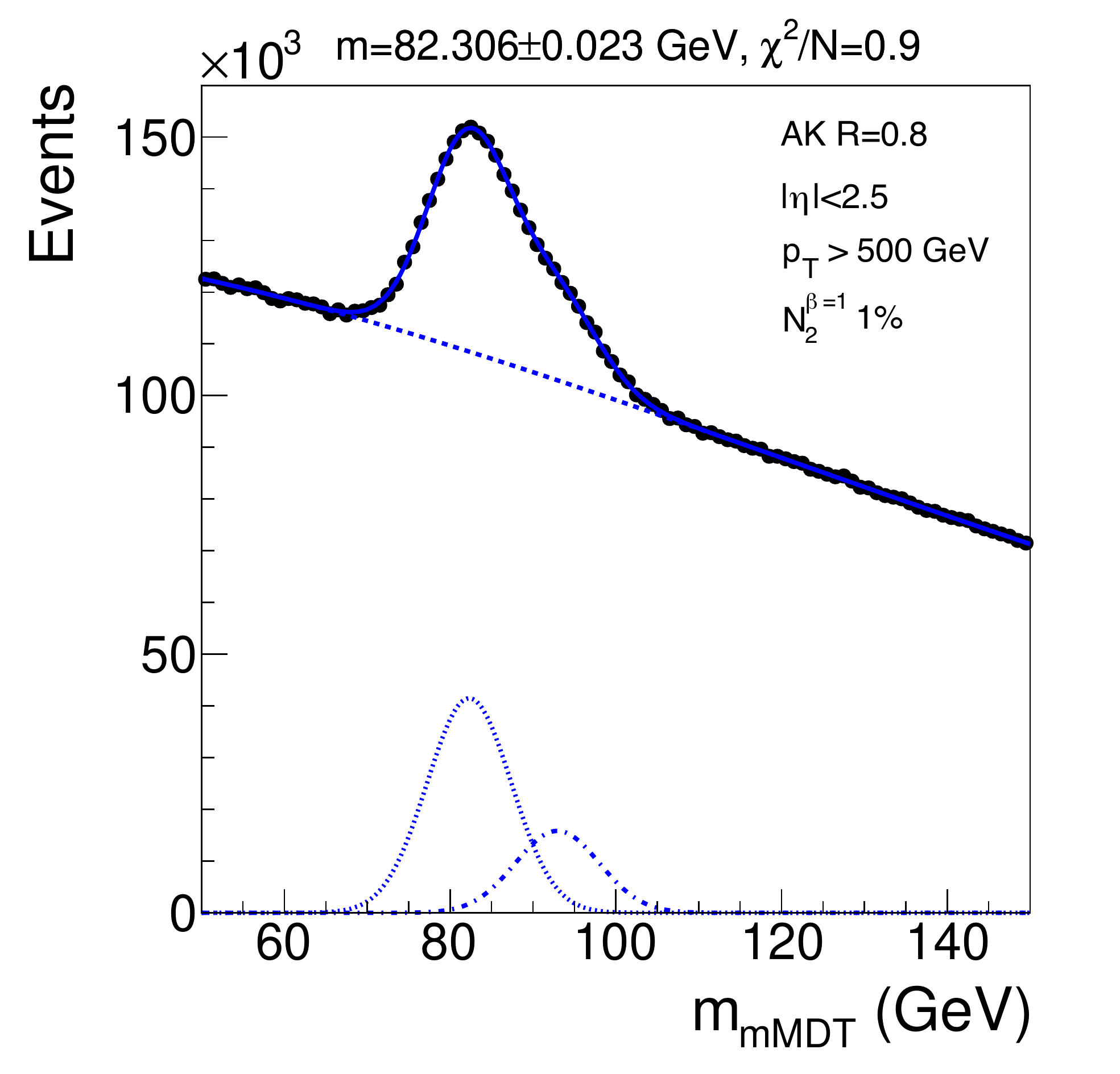}
\includegraphics[width=0.45\linewidth]{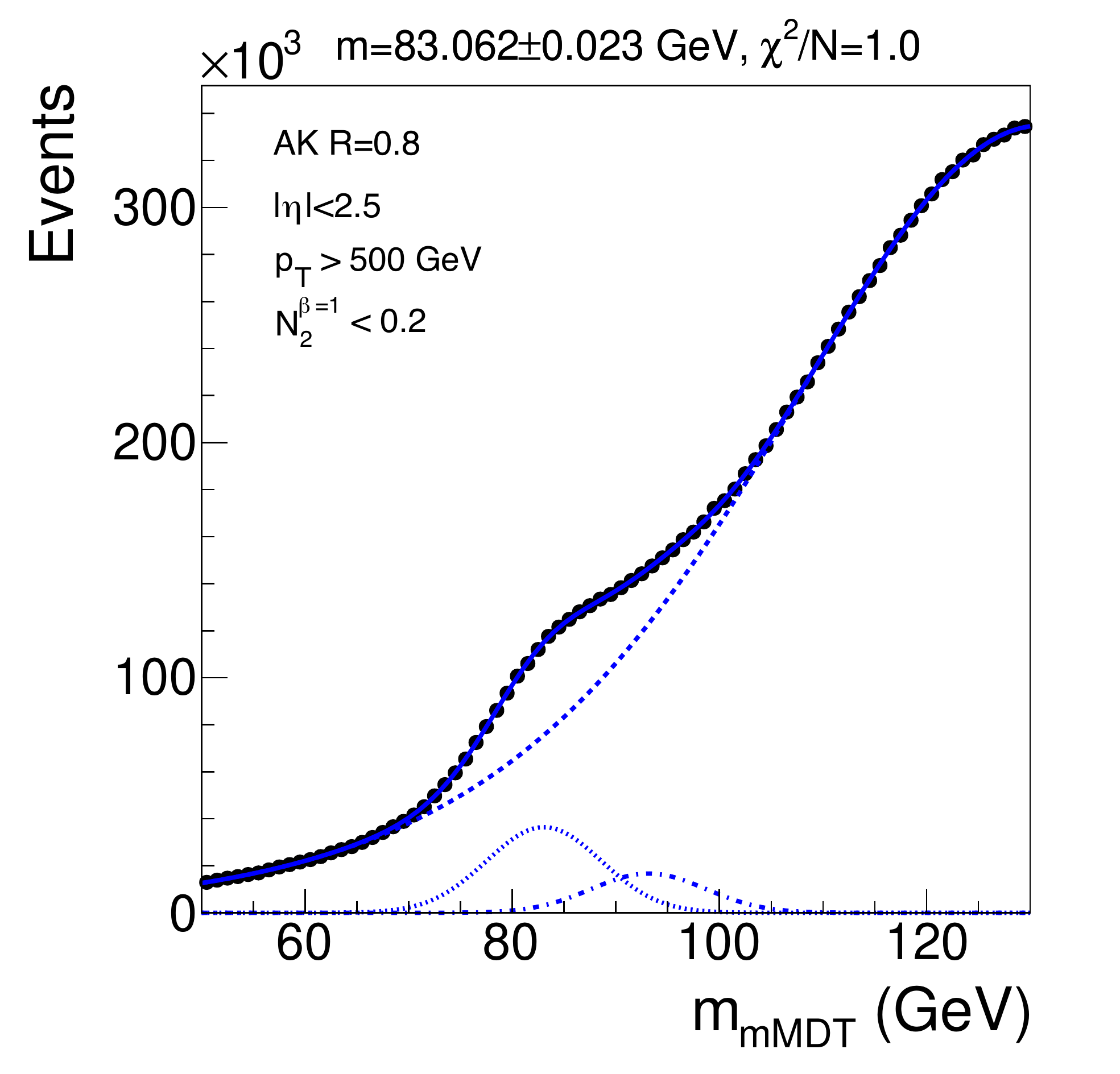}
\end{center}
\caption{Simultaneous fit of the $W$+$Z$ signal shape and the functional form parametrizing the background to a distribution of the mass $m_{\text{mMDT}}$ of the leading jet in the event from simulated signal and background pseudo-data corresponding to an integrated luminosity of 3000/fb.
The left plot requires a decorrelated N$_2^{\beta=1}$ 1\% tagger, the right plot is after a selection of N$_2^{\beta=1}<0.2$.
The fitted value of the $W$ mass peak $m_W$ and its statistical uncertainty is quoted on top of the figure together with the $\chi^2$ per degree of freedom, $N$, of the fit.}
\label{fig:fitWonly}
\end{figure}

We also study variants of $N_2^\beta$ that have been decorrelated~\cite{Dolen:2016kst} with jet \pt and $m_{\text{mMDT}}$ for QCD multijet events. Figure~\ref{fig:ddt} shows the correlation of $N_2^{\beta=1}$ with $\rho=\log(m_{\text{mMDT}}^2/\pt^2)$ for different bins of jet \pt in simulation of QCD multijet production, from which the 1\% quantile of the $N_2^{\beta=1}$ distribution for each \pt and $\rho$ bin is computed. The decorrelated $N_2^{\beta=1}$ 1\% tagger, denoted $N_2^{\beta=1}$ 1\%~$(\pt,\rho)$, requires $N_2^\beta$ to be smaller than this quantile, estimated from multijet simulation rather than a constant value across \pt and $m_\text{mMDT}$. From Fig.~\ref{fig:roc} it can be seen that the performance in terms of background rejection power vs.\ $W$ jet selection efficiency is similar to the taggers without decorrelation. The performance is also similar for different choices of the quantile of multijet reduction used in the decorrelation. However, by construction the decorrelated $N_2^{\beta=1}$ 1\% tagger guarantees a smoothly falling $m_{\text{mMDT}}$ spectrum for QCD multijet background for any $p_T$, simplifying the signal $W$ extraction procedure as demonstrated in Fig.~\ref{fig:fitWonly} (left) and explained in more detail in the next section.

\subsection{Measurement of $m_W$}

We first consider an approach to obtain the $W$ boson mass by measurement of the peak position of the $W$ boson \emph{alone} on top of the smoothly falling QCD mutlijet background, treating $Z$ boson production as a background.
Since top quark production can be measured in lepton+jet events, we assume its all-hadronic contribution can estimated precisely and subtracted, and thus do not consider it in this feasibility study, it contributes $\sim$5\% to the sample of $W$ bosons selected here.

The current lowest unprescaled trigger thresholds on the leading large radius jet $p_T$ collecting multijet events plateau at 100\% efficiency for a reconstructed jet \pt of 500~GeV for the CMS~\cite{Sirunyan:2017acf} and ATLAS experiments~\cite{Aaboud:2016leb}.
Due to the foreseen increase in luminosity, it will become challenging to preserve the online jet momentum selection at its current value until the end of LHC and HL-LHC running.
In the following, we assume that we will succeed in retaining events with the leading jet passing $\pt>500$ GeV and the decorrelated  {$N_2^{\beta=1}$ 1\%} tagger.

As demonstrated in Fig.~\ref{fig:fitWonly} (left), the leading jet mass $m_{\text{mMDT}}$ for multijet background after this selection can be described by a smooth functional form, enabling a signal plus background fit to extract the signal parameters.
The background is parametrized by a logistic function with 3 free parameters:
\begin{equation}
     \text{bkg}(m) = \frac{p_0}{1+e^{-(m-p_1)/p_2}}.
\end{equation}
The signal is parametrized by a Gaussian function to simulated signal samples.
To estimate the expected statistical uncertainty of a $W$ boson measurement, we first generate pseudo-data from the signal plus background functional forms for the expected number of events corresponding to 30 and 300/fb of integrated luminosity at the LHC and 3000/fb at the HL-LHC. With these pseudo-data distributions, we perform signal plus background fits. The result of a fit for the HL-LHC scenario is shown in Fig.~\ref{fig:fitWonly} (left).
The estimated statistical uncertainties for $W$ mass measurements at the LHC and HL-LHC are summarized in Table~\ref{stat_uncertainty_table1}.

\begin{table}[htb]
\begin{center}
\label{models}
\caption{\label{stat_uncertainty_table1}Statistical uncertainty of an $m_W$ mass measurement for different selections. }
\begin{tabular}{l|l|l|c}
Strategy & Selection & Int. luminosity & $\sigma_{m_W}$ [MeV]\\
\hline
measure $m_W$ & decorrelated $N_2^{\beta=1} 1\%$, $\pt>500$ GeV & 30/fb & 110 \\
measure $m_W$ & decorrelated $N_2^{\beta=1} 1\%$, $\pt>500$ GeV & 300/fb & 75 \\
measure $m_W$ & decorrelated $N_2^{\beta=1} 1\%$, $\pt>500$ GeV & 3000/fb & 23 \\
measure $m_W$ & $N_2^{\beta=1}<0.2$, $\pt>500$ GeV & 3000/fb & 23 \\
\end{tabular}
\end{center}
\end{table}

In Fig.~\ref{fig:fitWonly} (right), we demonstrate how the same approach performs without decorrelating $N_2^{\beta=1}$.
While a similar statistical uncertainty can be achieved as for the decorrelated tagger, the procedure is subject to larger systematic uncertainties due to the necessity for a background functional form that is not smoothly falling unlike the decorrelated tagger.
The differences of 756 MeV in peak position between the correlated and decorrelated tagger on top of the background can be taken as an indication of the size of background systematic effects without decorrelation.
Further, when lowering the $p_T$ threshold below 500~GeV, without decorrelating $N_2^{\beta=1}$, the maximum of the background jet mass spectrum would peak close to the $W$ mass, making the extraction of the $W$ boson peak even more challenging.

For each tagger, we have considered working points with 0.5\%, 1\%, 2\%, 5\% and 10\% quark and gluon jet efficiency, jet mass ranges from 20 to 200 GeV and various background functional forms.
We quote in Table~\ref{stat_uncertainty_table1} the results that minimize the statistical uncertainty on the $W$ boson mass, while maintaining a good fit of the background functional form to QCD multijet simulation.
We have also studied an alternative decorrelated observable $N_2^{\beta=2}$ 2\%, which yields a slightly worse statistical uncertainty, but is found to be within 30\% of the best variable. This result is representative of the variance over the taggers considered in this study.

\subsection{Measurement of $\Delta m = m_Z-m_W$}

The above approach to measure exclusively the $W$ mass is highly sensitive to uncertainties related to the absolute calibration of the jet mass. We consider an alternative approach where the $Z$ boson mass peak is used as a standard candle to constrain experimental uncertainties related to the jet mass calibration and theoretical uncertainties related to the jet mass spectrum prediction.

\begin{figure}[htb]
\begin{center}
\includegraphics[width=0.45\linewidth]{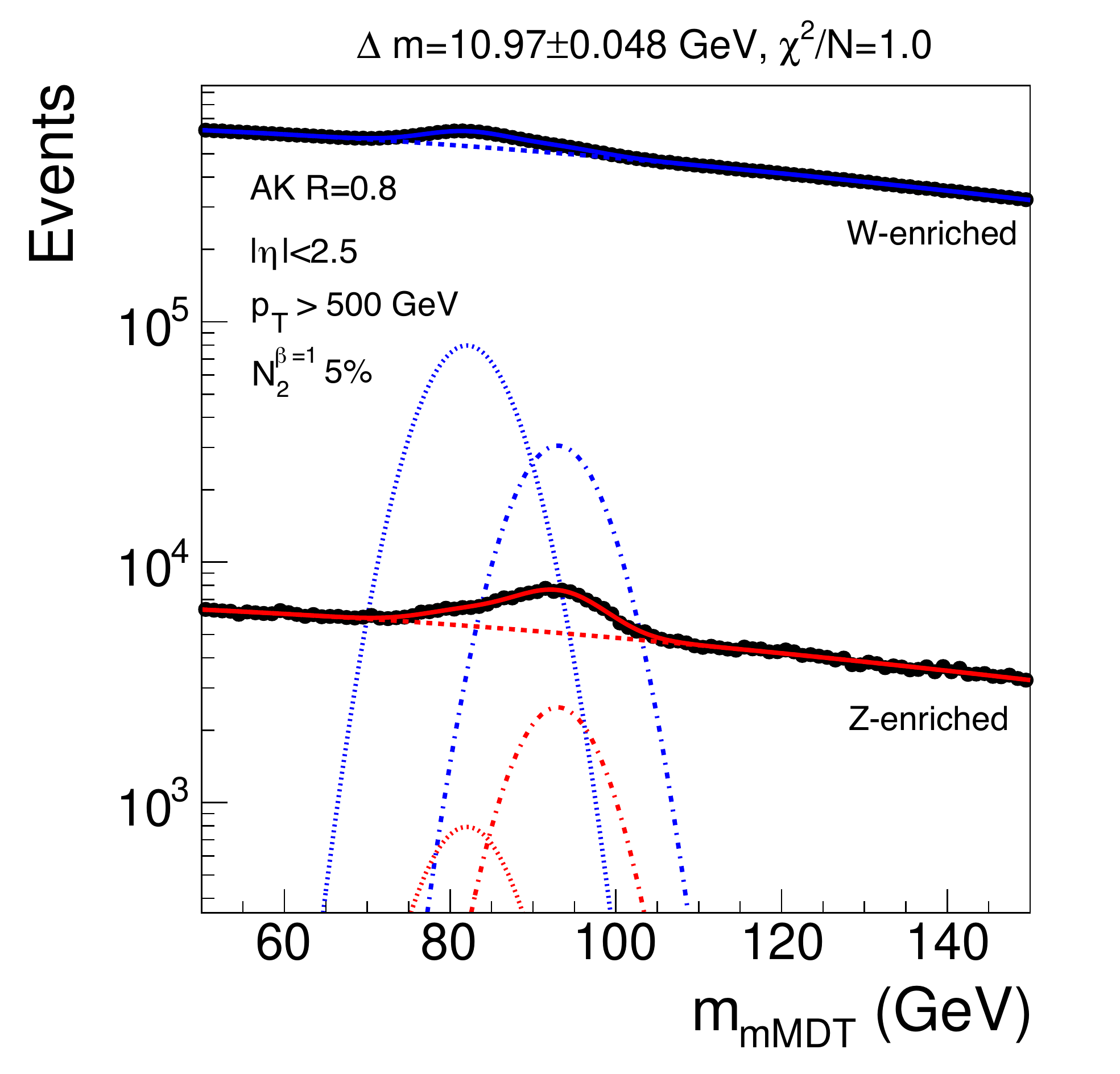}
\end{center}
\caption{Simultaneous fit of the $W$+$Z$ signal shape plus the functional form parametrizing the background to two distributions of the mass $m_{\text{mMDT}}$ of the leading jet in the $Z$-enriched and $W$-enriched event categories from simulated signal and background pseudo-data corresponding to an integrated luminosity of 3000/fb.
The fitted value of the mass difference between the $Z$ peak and the $W$ peak $\Delta m = m_Z-m_W$ and its statistical uncertainty is quoted on top of the figure together with the $\chi^2$ per degree of freedom, $N$, of the fit.}
\label{fig:fitZW}
\end{figure}

The data sample is split into a category enriched with $Z$ bosons and a category enriched in $W$ bosons by using a $b$-tagging algorithm to exploit the higher branching fraction to $b$ quarks (15.12\% for $b\bar{b}$) to that of the $W$ boson (0.06\% to $b\bar{c}$). 
A $b$-tagger makes use of the fact that $b$ quarks form $B$ hadrons that have a larger lifetime than lighter hadrons and can be identified by secondary decay vertices made of tracks with large impact parameters with respect to the primary vertex and several observables that characterize $B$ hadrons flight directions in relation to the jet substructure.
We consider an efficiency of 45\% for $Z \to b\bar{b}$ jets and an efficiency of 1\% for light quark jets and 3\% for $Z \to c\bar{c}$ jets.
These efficiency values are similar to the performance of double $b$-taggers employed by both the CMS~\cite{CMS-PAS-BTV-15-002} and ATLAS~\cite{ATLAS-CONF-2016-039} experiments for large radius jets.
Fig.~\ref{fig:fitZW} shows the resulting expected $W$ and $Z$ signal peaks in the $Z$-enriched and the $W$-enriched categories. By measuring the mass difference no absolute calibration of the jet mass is needed, instead only the relative jet mass scale of light and $b$-flavor enriched jets need to be calibrated.

The result of the simultaneous fit to the two categories to pseudo data for the HL-LHC is shown in Fig.~\ref{fig:fitZW}.
The statistical uncertainty of the $m_Z-m_W$ mass measurement at the LHC and HL-LHC is given in Table~\ref{stat_uncertainty_table2}.
The statistical uncertainty is larger than for the case of the $m_W$ mass measurement, because the sample of $Z$ bosons decaying to $b$ quarks is significantly smaller than that of $W$ bosons decaying to quarks.
We have considered double $b$-tagger selections corresponding to 10, 30, 45, 75, and 85\% $Z \to b\bar{b}$ jet efficiency to divide up the data into $Z \to b\bar{b}$-enriched and the $W \to q\bar{q}$-enriched categories. We report the result for the selection corresponding to 45\% $Z \to b\bar{b}$ jet efficiency that yields the best statistical precision for the $m_Z-m_W$ mass measurement.
However, it should be noted that yet better precision may be achievable by splitting the sample further into a single-$b$-enriched and double-$b$-enriched and light quark enriched samples.
Correlating or not correlating the background shape parameters between the two categories does not change the resulting statistical uncertainty on the $m_Z-m_W$ mass measurement, which indicates that the measurement uncertainty is mainly driven by the $W$ and $Z$ peak modeling rather than modeling of the smoothly falling background shape.

\begin{table}[tb]
  \begin{center}
    \begin{tabular}{l|l|l|c}
      Strategy & Selection & Int. luminosity & $\sigma_{m_W}$ [MeV]\\
      \hline
      measure $m_Z-m_W$ & decorrelated $N_2^{\beta=1} 2\%$, $\pt>500$ GeV & 30/fb & 500 \\
      measure $m_Z-m_W$ & decorrelated $N_2^{\beta=1} 2\%$, $\pt>500$ GeV & 300/fb & 171\\
      measure $m_Z-m_W$ & decorrelated $N_2^{\beta=1} 5\%$, $\pt>500$ GeV & 3000/fb & 48 \\
    \end{tabular}
  \end{center}
  \caption{Statistical uncertainty of $\Delta m = m_Z-m_W$ mass measurement for different selections.}
  \label{stat_uncertainty_table2}
\end{table}

\subsection{Measurement with New Triggers}

Following existing triggers, the best estimated statistical uncertainty from both approaches at the HL-LHC is a factor 2 worse than the precision reached by existing measurements of the $W$ boson mass.
Though a moderate improvement is expected combining this measurement with the existing ones due to largely uncorrelated uncertainties, we propose to make use of new trigger strategies for this measurement to increase the number of $W\to q\bar{q}'$ events, following the data scouting approaches explored by CMS~\cite{Khachatryan:2016ecr}, ATLAS~\cite{Aad:2014aqa} and LHCb~\cite{Benson:2015yzo}, where only limited event information is stored to allow data storage at a higher rate yielding a lower high-level trigger jet $p_T$ threshold.
The data rate after high-level trigger is the limiting factor for the jet $p_T$ threshold at the moment and will likely remain so even at the HL-LHC, because hardware-based triggers using jet substructure information~\cite{Collaboration:2283192} may become feasible to maintain low hardware-based jet $p_T$ trigger thresholds.
Though trigger level jets currently have larger associated systematic uncertainties than offline jets, we will assume here that the advances in the trigger system for the HL-LHC will allow to achieve the same systematic uncertainties as for offline jets.

\begin{table}[tb]
  \begin{center}
    \begin{tabular}{l|l|l|c}
      Strategy & Selection & Int. luminosity & $\sigma_{m_W}$ [MeV]\\
      \hline
      measure $m_W$ & decorrelated N$_2^{\beta=1} 2\%$, $\pt>300$ GeV & 300/fb & 40 \\
      measure $m_W$ & decorrelated N$_2^{\beta=1} 1\%$, $\pt>500$ GeV & 3000/fb & 23 \\
      measure $m_W$ & decorrelated N$_2^{\beta=1} 1\%$, $\pt>400$ GeV & 3000/fb & 21\\
      measure $m_W$ & decorrelated N$_2^{\beta=1} 2\%$, $\pt>300$ GeV & 3000/fb & 13 \\
      \hline
      measure $m_Z-m_W$ & decorrelated N$_2^{\beta=1} 2\%$, $\pt>300$ GeV & 300/fb & 99\\
      measure $m_Z-m_W$ & decorrelated N$_2^{\beta=1} 5\%$, $\pt>500$ GeV & 3000/fb & 48\\
      measure $m_Z-m_W$ & decorrelated N$_2^{\beta=1} 5\%$, $\pt>400$ GeV & 3000/fb & 40\\
      measure $m_Z-m_W$ & decorrelated N$_2^{\beta=1} 5\%$, $\pt>300$ GeV & 3000/fb & 32\\
    \end{tabular}
  \end{center}
  \caption{Statistical uncertainty of $W$ mass measurement for different strategies and trigger selections.
  The $\pt>300$ and $\pt>400$ selections will require new trigger strategies.}
  \label{stat_uncertainty_table3}
\end{table}

Table~\ref{stat_uncertainty_table3} presents scenarios with jet \pt trigger thresholds lowered to $\pt>300$ GeV and $p_T>400$ GeV at the LHC and HL-LHC.
The statistical uncertainties are significantly reduced with lower trigger thresholds, though the achievable statistical uncertainty with the integrated luminosity of 3000/fb and $\pt>500$ GeV still remains lower than that of 300/fb and $\pt>300$ GeV.
The ultimate statistical uncertainty on the $W$ boson mass measurement that can be achieved is 13 MeV for the $m_W$ approach and 32 MeV for the $\Delta m = m_Z-m_W$ approach.

\section{Systematic Uncertainties}
\label{sec:uncertainties}

Rather than estimating the uncertainty based on the current knowledge of LHC detectors and theory~\cite{Aaboud:2017svj}, we provide an estimate of the experimental and theoretical precision in various sources of uncertainty that would be needed to achieve a systematic uncertainty of $\sigma_{m_W}=10$ MeV in a $W$ mass measurement.
The signal to background ratio and the continuity of the background distribution when selecting events with a decorrelated substructure observable are well sufficient to unambiguously separate the signal contribution from background. Uncertainties in the modeling of backgrounds are therefore not discussed, as their contribution to the mass measurement is expected to be subdominant.
Unless stated otherwise, a selection of $\pt > 300$ GeV and $N_2^{\beta=1}<0.2$ is applied when studying the signal systematic uncertainties, where no decorrelation is applied to $N_2^{\beta=1}$ to ease comparison to (future) theoretical computations of the observable.

\subsection{Experimental Uncertainties}

If the $W$ mass is measured without the use of $b$-tagging and the $Z$ mass peak, the jet mass needs to be calibrated with high precision. In Table~\ref{sys_effect_table}, we quantify what precision that is needed for the energy scale measurement of charged particles, photons (and $\pi^0$), and neutral hadrons. We quote the precision for each particle type, assuming perfect description of the other particle types.
The necessary precision for charged particles of 0.03\% and photons of 0.06\% is within a factor of 2 of what is currently achieved by the CMS~\cite{Sirunyan:2017ulk} and ATLAS~\cite{Aaboud:2017aca} detectors. The precision needed for neutral hadrons of 0.1\% is however an order of magnitude better than what is currently achieved e.g., in jet calibration~\cite{Khachatryan:2016kdb,Aaboud:2016hwh}. For an overall precision of $\sigma_{m_W}$$=$$10$ MeV, each particle type needs to be calibrated such that $\sqrt{\sigma_{m_{W,\text{charged particles}}}^2+\sigma_{m_{W,\text{photons}}}^2+\sigma_{m_{W,\text{neutral hadrons}}}^2}<10$ MeV. Unless this precision can be achieved with the large HL-LHC dataset, the measurement of $m_{Z}-m_{W}$ using $b$-tagging and the $Z$ mass peak would be the only feasible approach, although consequent improvements to generic boosted light resonance searches are not tied to such a specific benchmark, as discussed in the introduction.

\begin{table}[tb!]
  \begin{center}
    \begin{tabular}{l|l|c|c}
      Quantity & Effect & Understanding needed          & Typical current\\
               &        & for $\sigma_{m_W}$$=$$10$ MeV & precision \\
      \hline
      $m_W$ & Charged particle energy scale & 0.03\% & $0.05$\% \\
      $m_W$ & Photon (and $\pi^0$) energy scale & 0.06\% & $0.1$\% \\
      $m_W$ & Neutral hadron energy scale & 0.1\% & $1$\% \\
      $m_W$ & 200 pileup interactions & 1.4\% & $1$\% \\
      $m_Z$ & $Z\to q\bar{q}$ vs. $Z\to b\bar{b}$ & 7\% & $0.5$\% \\
    \end{tabular}
    \caption{List of systematic effects. Unless stated otherwise a selection of $\pt > 300$ GeV and $N_2^{\beta=1}<0.2$ is applied. Experimental systematic uncertainties are estimated using detector simulations.}
    \label{sys_effect_table}
  \end{center}
\end{table}

For a measurement of $m_Z-m_W$, an additional uncertainty arises from the understanding of the difference in detector response for $b$-, $c$-, and light quark-initiated showers. The effects of hadronization on the $W \to q\bar{q}'$ and $Z \to b\bar{b}$ mass distributions will be discussed in Sec.~\ref{sec:theory} and will need an improvement in understanding to a 5--10\% level for a precision of 10 MeV in $m_Z-m_W$. Previous measurements of the difference between $b$ and light jet energy response using a $Z+ \text{$b$-jet}$ balancing method achieved a precision of 0.5\%~\cite{Sirunyan:2017ulk}, thus well below the corresponding theoretical uncertainty

Another important experimental effect comes from additional $pp$ interactions happening in the same bunch crossing, so-called pileup interactions. Particles from pileup interactions enter the reconstructed jets of the main interaction and increase their mass. At the HL-LHC, up to 200 of such interactions can happen simultaneously. Dedicated suppression techniques have been developed to remove such contamination, among which we exploit the pileup per particle identification (PUPPI) algorithm~\cite{Bertolini:2014bba}. We estimate the shift in jet mass expected with 200 pileup interactions, applying the PUPPI algorithm, and quote how precise the pileup fractional estimate would be needed to model this on average.
To reach a $W$ mass precision of 10 MeV, the modeling of the extra jet mass from 200 pileup interactions needs to be at the level of 1.4\% (see Table~\ref{sys_effect_table}), which may seem feasible given the achieved 1\%-level modeling of the fraction of energy from pileup interactions in jets with $p_T>300$ GeV~\cite{Sirunyan:2017ulk}.
Maintaining this level of the understanding of pileup will, however, remain to be seen.
Both ATLAS and CMS experiments plan to upgrade their detection capabilities by introducing better tracking and timing detection systems.
The suppression and modeling of up to 200 pileup interactions will nevertheless remain a major experimental challenge.

\subsection{Theoretical Uncertainties}
\label{sec:theory}

\begin{figure}[htb]
\begin{center}
\includegraphics[width=0.45\linewidth]{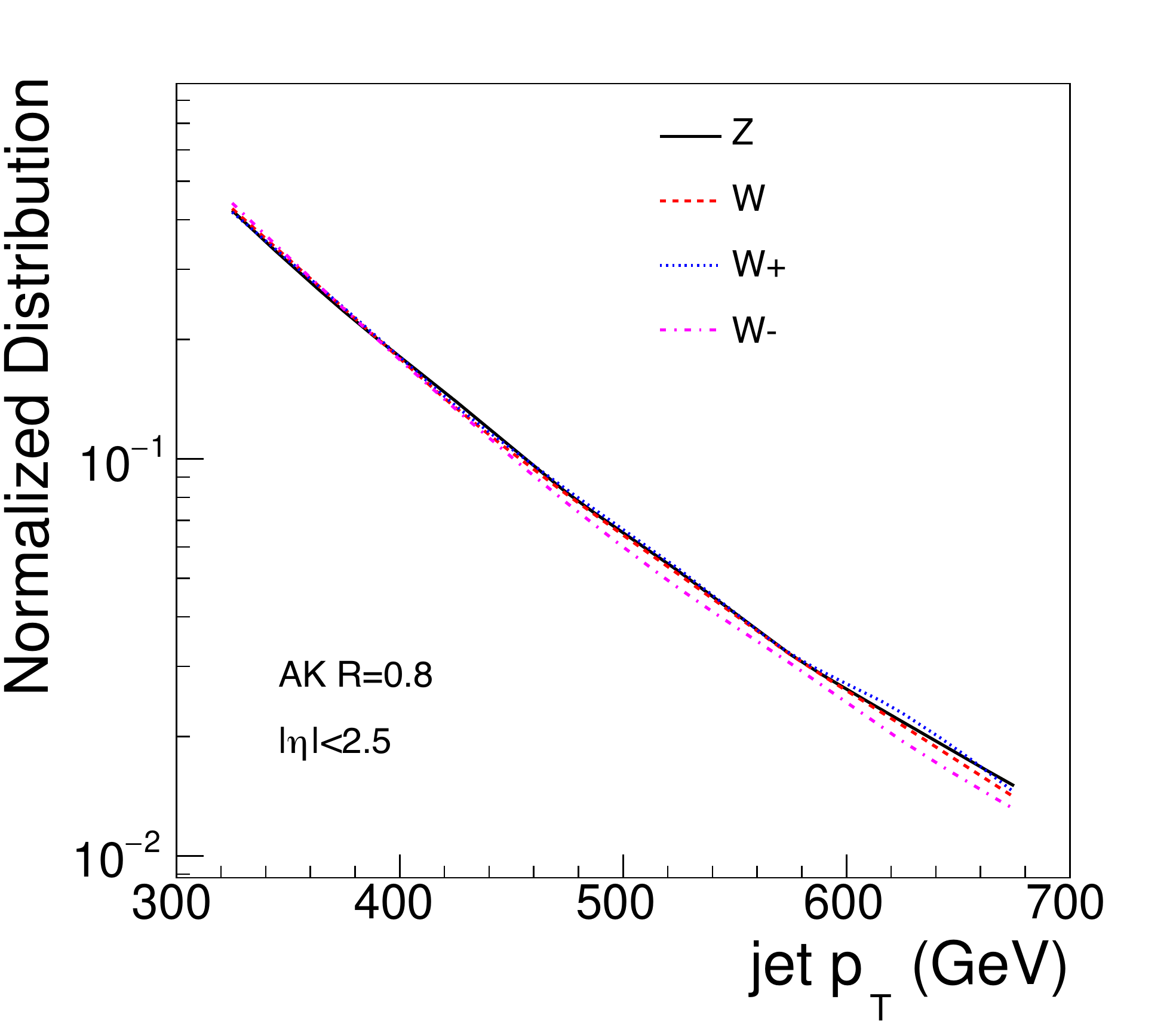}
\includegraphics[width=0.45\linewidth]{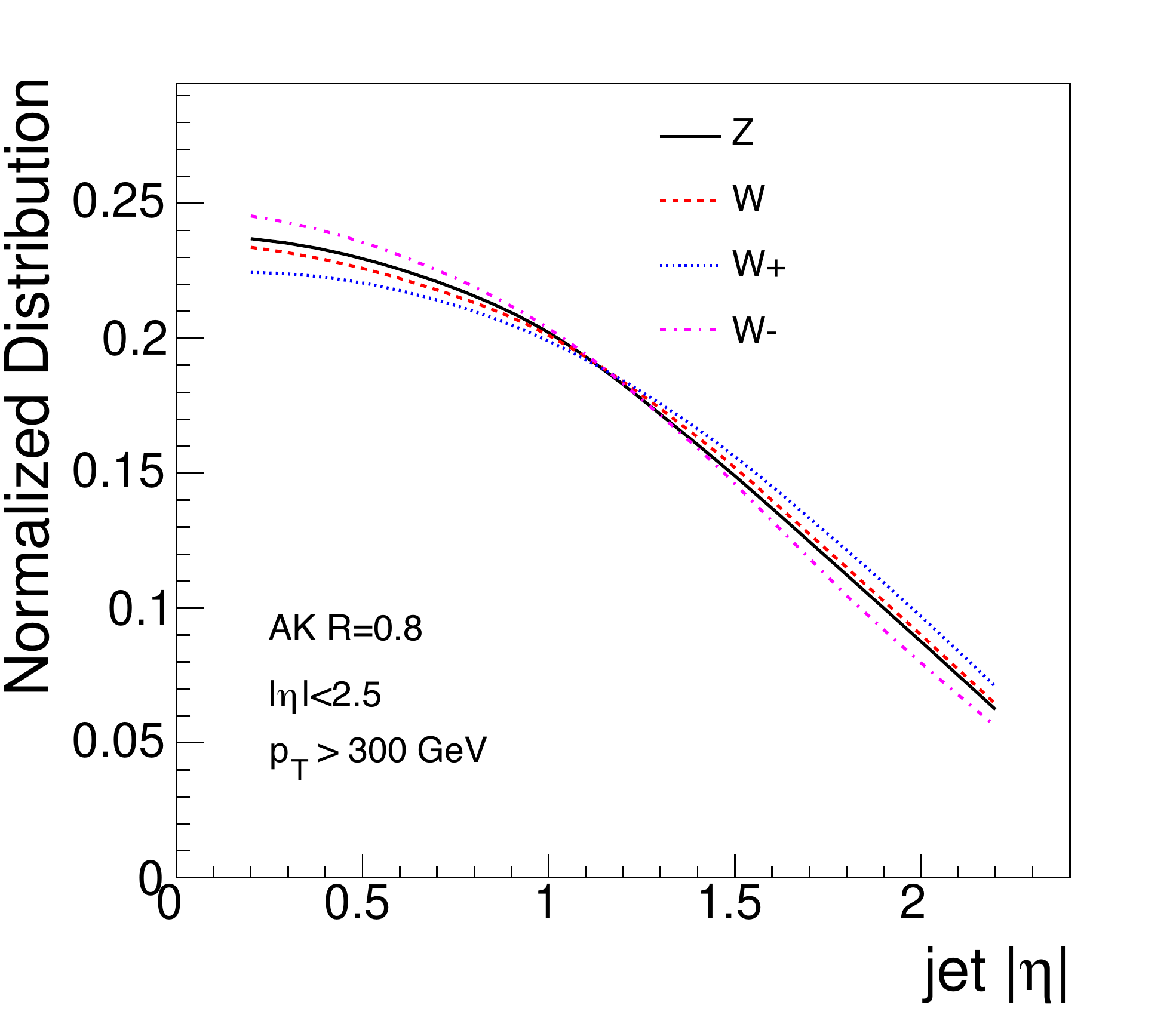}
\end{center}
\caption{Transverse momentum \pt and $|\eta|$ of the leading jet with $\pt>300$ GeV in particle level simulation of $W + \text{jets}$, $W^+ + \text{jets}$, $W^- + \text{jets}$ and $Z+ \text{jets}$.}
\label{fig:Wkinematics}
\end{figure}

\paragraph*{\boldmath $W$ and $Z$ Boson Kinematics:}
In order to extract the $W$ boson mass from the measured jet mass, sufficient theoretical precision in the prediction of the jet mass of a $W$ boson is also required.
Figure~\ref{fig:Wkinematics} shows the kinematic distributions of the leading jet in $W$ and $Z$ production.
The jet \pt, mass and $\eta$ distributions are subject to multiple theoretical uncertainties, including parton density functions, factorization and renormalization scales, hadronization models.
To demonstrate the dependence of the $W$ jet mass on the prediction of the jet kinematics, we compute the mass difference between $W^+$ and $W^-$ jets to 170 MeV, which have different $p_T$ and $\eta$ distributions due to the parton composition of the colliding protons.
The uncertainty on the $W$ mass due to parton density functions, factorization and renormalization scales used for the prediction of the \pt spectra of the $W$ and $Z$ is evaluated by reweighing them to NNLO predictions and varying them according to the NLO PDF, NLO QCD and NLO EW uncertainties computed in Ref.~\cite{Lindert:2017olm}.
The resulting uncertainties quoted in Table~\ref{sys_effect_table2} show that current predictions of \pt spectra for the $W$ and $Z$ have sufficient precision for a 10 MeV $W$ mass measurement, considering only these uncertainty sources.
When taking into account that other experimental or theoretical uncertainties of similar size can contribute, small improvements over the current precision is desirable.

\begin{figure}[htb]
\begin{center}
\includegraphics[width=0.45\linewidth]{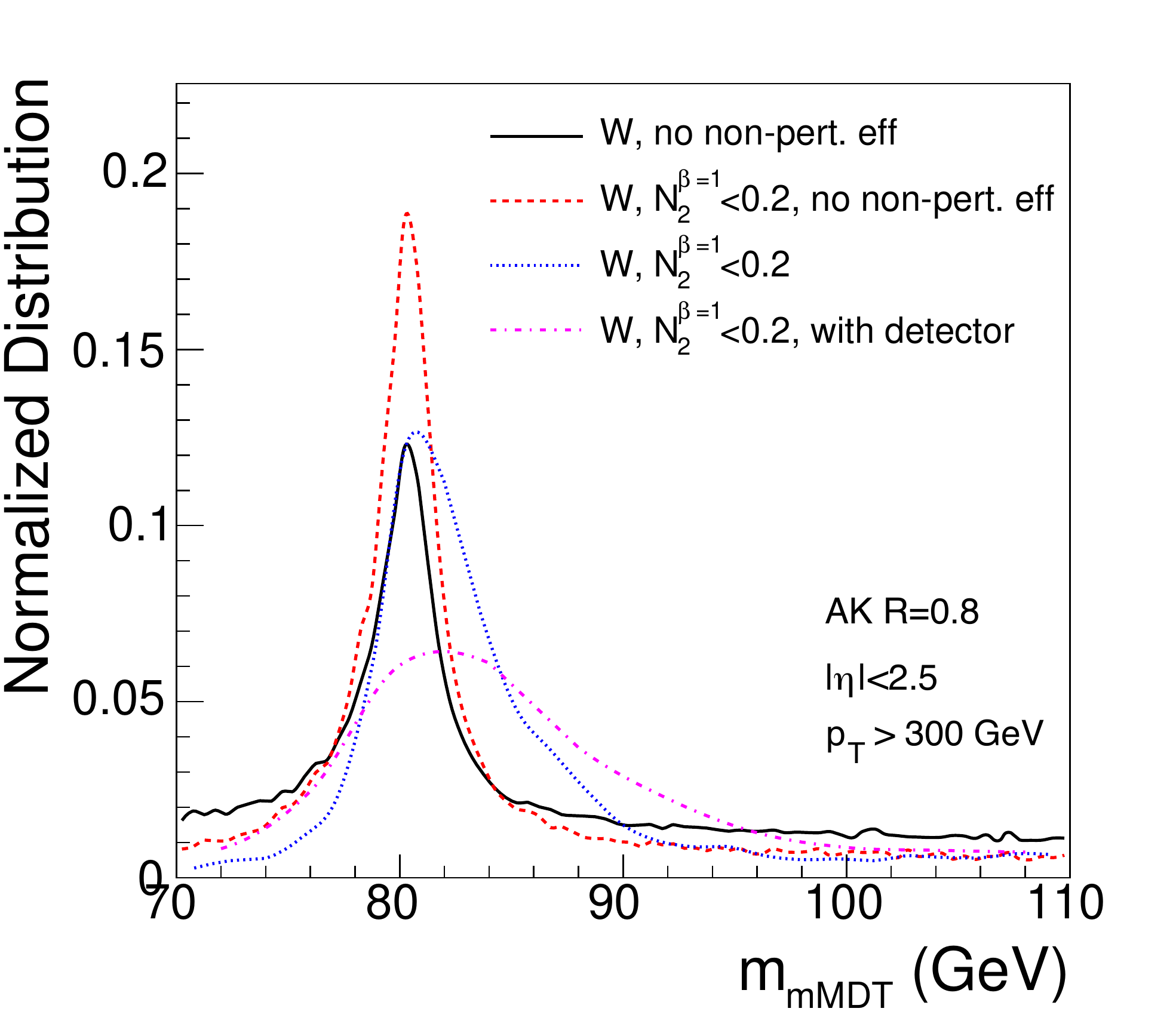}
\end{center}
\caption{Dependence of the $W$ jet mass distribution on non-perturbative corrections, $N_2^{\beta=1}$ selection and detector simulation.}
\label{fig:effects}
\end{figure}

\begin{table}[htb!]
\begin{center}
\label{models}
\begin{tabular}{l|l|c|c}
Quantity & Effect & Size of effect & Understanding needed\\
 &  & & for $\sigma_{m_W}$$=$$10$ MeV\\
\hline
$m_W$ & NLO QCD & 8 MeV & \checkmark \\
$m_W$ & NLO EW & 1 MeV & \checkmark \\
$m_W$ & NLO PDF & 1 MeV & \checkmark \\
$m_W$ & $N_2^{\beta=1}<0.2$ selection & 310 MeV & 3\% \\
$m_W$ & non-pert.\ corrections & 1100 MeV & 0.9\% \\
$m_W$ & $W\to q\bar{q}'$ vs. $W\to c\bar{s}$ & 80 MeV & 13\% \\
\hline
$m_Z-m_W$ & NLO QCD & 4 MeV & \checkmark \\
$m_Z-m_W$ & NLO EW & 1 MeV & \checkmark \\
$m_Z-m_W$ & NLO PDF & 1 MeV & \checkmark \\
$m_Z-m_W$ & $N_2^{\beta=1}<0.2$ selection & 200 MeV & 5\% \\
$m_Z-m_W$ & non-pert.\ corrections & 110 MeV & 9\% \\
$m_Z$ & $Z\to q\bar{q}$ vs. $Z\to b\bar{b}$ & 140 MeV & 7\% \\
\end{tabular}
\caption{\label{sys_effect_table2}List of systematic effects. The understanding needed for $\sigma_{m_W}=10$ MeV is the fraction of 10 MeV and the quoted size of effect. It should be noted that yet better precision is needed to achieve a sum in quadrature of all systematic uncertainties of $\sigma_{m_W} = 10$ MeV. Unless stated otherwise a selection of $\pt > 300$ GeV and $N_2^{\beta=1}<0.2$ is applied. Theoretical systematic uncertainties are estimated using particle-level simulations.}
\end{center}
\end{table}

\paragraph*{Non-perturbative Effects and Jet Substructure:}

Additionally, the non-perturbative effects and the impact of jet substructure selection on the $W$ jet mass must be theoretically understood.
Figure~\ref{fig:effects} shows the dependence of the $W$ jet mass distribution on various effects.
The non-perturbative effects are studied by disabling hadronization and underlying event in \textsc{Pythia} and computing the jet mass from partons after showering. As a cross check, we also consider the difference between \textsc{Herwig} and \textsc{Pythia}, which are based on different models for non-perturbative effects and parton showering. Excluding non-perturbative effects, a significant difference between the jet mass with and without a jet substructure selection of $N_2^{\beta=1}<0.2$ is observed and quoted in Table~\ref{sys_effect_table2}. For a $W$ mass measurement, it would require an understanding of the observed difference at a 3\%-level for a $W$ mass precision of 10 MeV. The $W$ jet mass distribution with and without non-perturbative effects after $N_2^{\beta=1}<0.2$ selection shows a difference of similar order, that would require an understanding of this difference at a 9\% (0.9\%)-level for a $W$ mass precision of 10 MeV with (without) the use of the $Z$ mass peak. Other grooming choices may further reduce these uncertainties. The difference between $m_Z-m_W$ ($m_W$) in \textsc{Herwig} and \textsc{Pythia} after $N_2^{\beta=1}<0.2$ selection is also of similar order, ranging from 50--500 MeV (200--1000 MeV) depending on the grooming algorithm used. It would thus require an understanding of their difference at 2--20\% (1--5\%)-level for a $W$ mass precision of 10 MeV with (without) the use of the $Z$ mass peak. In all cases, a significant improvement in understanding of these non-perturbative and showering effects needs to be reached to make this measurement feasible. One should also note that the $m_\text{mMDT}$ mass distribution with non-perturbative effects as shown in Fig.~\ref{fig:effects} is no longer similar to a Gaussian distribution, but rather asymmetric, which may complicate the definition of the mass peak position. This jet mass distribution shape depends strongly on the choice of grooming algorithm and differs between \textsc{herwig} and \textsc{pythia}. Appendix~\ref{sec:appendix} shows examples of such distributions. We find the jet mass distribution becomes increasingly more asymmetric when going from $\pt > 500$ GeV to $\pt > 300$ GeV. At a jet $\pt > 500$ GeV the difference between $m_Z-m_W$ \textsc{Herwig} and \textsc{Pythia} is strongly reduced from that of lower \pt jets to 23 MeV for mMDT. This variation ranges from 10-50 MeV for different groomers. By making use of the $p_T$  dependence of the jet mass, there is thus potential to constrain the contributions from non-perturbative effects.
Thinking beyond the HL-LHC, one should note that this measurement will also benefit from higher center of mass energies at future colliders (e.g. HE-LHC~\cite{Todesco:2011np} or FCC~\cite{Arkani-Hamed:2015vfh}) due to the fact that non-perturbative effects are significantly reduced for higher W boson \pt.

\begin{figure}[htb]
\begin{center}
\includegraphics[width=0.45\linewidth]{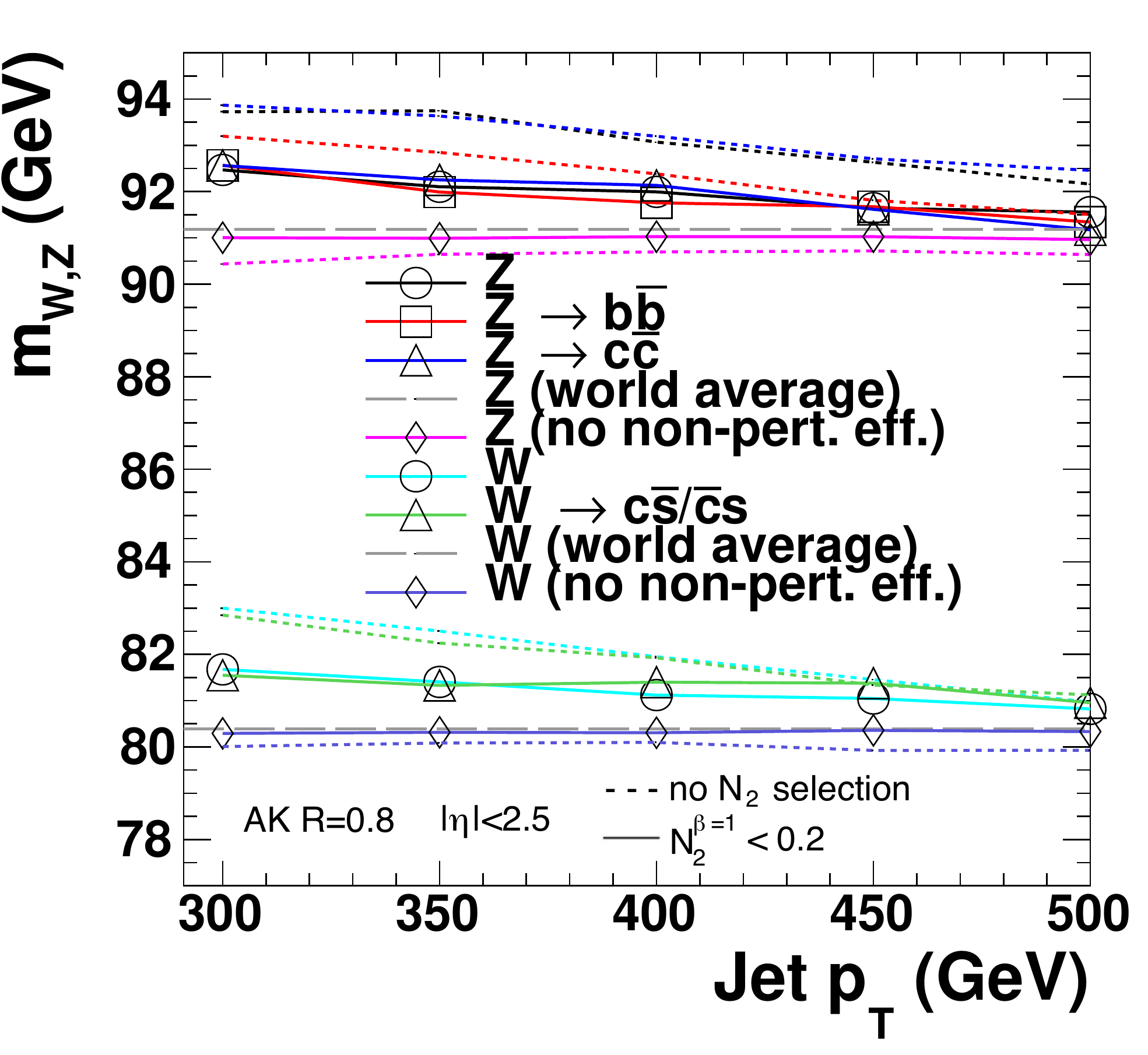}
\includegraphics[width=0.45\linewidth]{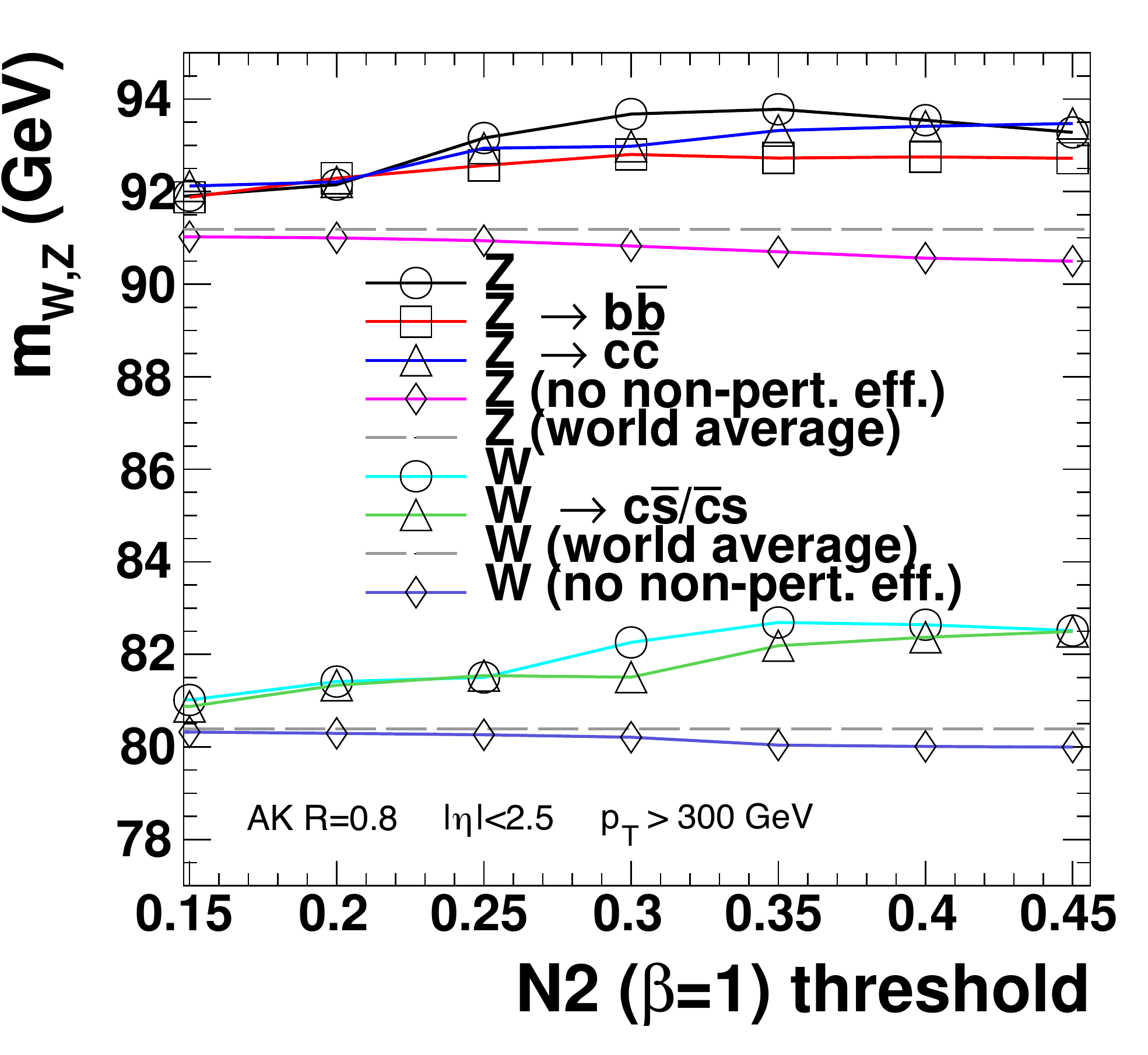}
\end{center}
\caption{Dependence of the $W$ and $Z$ jet mass on the jet $p_T$ and N$_2^{\beta=1}$ selection in particle-level simulation.
In the left plot solid lines correspond to a selection of $N_2^{\beta=1}<0.2$ and dotted lines to no $N_2^{\beta=1}$ selection. Curves labeled ``no non-pert.\ eff.'' correspond to predictions without simulation of hadronization and underlying event effects. The measured world average values of the $W$ and $Z$ mass taken from Ref.~\cite{Patrignani:2016xqp} are shown for comparison.}
\label{fig:dependence}
\end{figure}

Figure~\ref{fig:dependence} shows the dependence of the $W$ and $Z$ jet mass on jet \pt and $N_2^{\beta=1}$ selection.
The variation of the $W$ and $Z$ jet mass as a function of \pt is minimal with $N_2^{\beta=1}$ selection applied and no non-perturbative effects considered.
When non-perturbative effects are taken into account, the dependence of the $W$ and $Z$ jet mass on \pt are enhanced.
However, since $W$ and $Z$ show similar trends as a function of \pt, the contribution from non-perturbative effects is reduced for a $m_Z - m_W$ measurement, as can also be seen in Table~\ref{sys_effect_table2}.
Similarly, the $N_2^{\beta=1}$ dependence plot shows the largest shifts when adding non-perturbative effects.
However, a similar trend is observed between $W$ and $Z$ bosons.
For a very loose selection on $N_2^{\beta=1}$ a small deviation in perturbative effects is also observed.  

\paragraph*{\boldmath $b$-quark Hadronization:}

When considering a $W$ mass measurement using the $Z$ mass peak, an additional uncertainty arises from the theoretical knowledge of the jet mass difference at particle level between the $W$ and $Z$ peak. In particular, the hadronization of $b$ quarks will result in larger uncertainties in the $b$-hadron decay due to the possible presence of neutrinos in the final state. We quantify the impact of this modeling by computing the difference between the jet mass arising from $Z\to q\bar{q}$ and $Z\to b\bar{b}$ decays in Table~\ref{sys_effect_table2}. The addition of jet substructure selection significantly reduces this difference. The measurement thus benefits from the jet substructure selection, not only through reduction of background, but also through reduction of systematic effects.

To cross check the size of this effect, we compute the difference between \textsc{Herwig} and \textsc{Pythia} for the difference between $Z\to q\bar{q}$ and $Z\to b\bar{b}$ jet masses for different grooming algorithms. This difference would need an understanding in the 2--20\% range for a precision of 10 MeV, indicating a significant improvement in understanding non-perturbative and showering effects on $b$-hadronization would be needed to make this measurement feasible. Simultaneously, in \textsc{Pythia} lone, $W$ mass precision of 10 MeV for jets with substructure selection can be achieved with understanding of the hadronization of $b$ quarks at the 7\% level, comparable to current $b$-hadronization uncertainties in dedicated analyses~\cite{DELPHI:2011aa}, indicative of heavy flavor fragmentation tuning between \textsc{Herwig} and \textsc{Pythia} currently having room for improvement. For $c$ quarks only a 13\%-level understanding of the difference between $W \to q\bar{q}'$ and $W \to c\bar{s}$ is needed as quoted in Table~\ref{sys_effect_table2}. The contribution of $Z \to c\bar{c}$ is suppressed by more than a factor 10 compared to $Z \to b\bar{b}$ with a typical double $b$-tagger.

\begin{table}[tb!]
  \begin{center}
    \begin{tabular}{l|c|c}
      Effect & Understanding needed          & Typical current\\
                & for $\sigma_{m_W}$$=$$10$ MeV & precision \\
      \hline
      200 pileup interactions & 1.4\% & $1$\% \\
      $Z\to q\bar{q}$ vs. $Z\to b\bar{b}$ & 7\% & $0.5$\% \\
      NLO QCD & \checkmark & 4 MeV \\
      NLO EW & \checkmark & 1 MeV \\
      NLO PDF & \checkmark & 1 MeV \\
      $N_2^{\beta=1}<0.2$ selection & 5\% & 200 MeV \\
      Non-pert.\ corrections & 9\% & 110 MeV \\
      \hline
      Statistics with 3000/fb & 32 MeV & 500 MeV \\
    \end{tabular}
    \caption{Summary of uncertainties for an $m_Z-m_W$ measurement. The understanding needed for $\sigma_{m_W} = 10$ MeV is the fraction of 10 MeV and the estimated size of effect. It should be noted that yet better precision is needed to achieve a sum in quadrature of all systematic uncertainties of $\sigma_{m_W} = 10$ MeV. Unless stated otherwise a selection of $\pt > 300$ GeV and $N_2^{\beta=1}<0.2$ is applied. Theoretical systematic uncertainties are estimated using particle-level simulations.}
    \label{summary_effect_table}
  \end{center}
\end{table}

In Table~\ref{summary_effect_table}, we summarize the most important uncertainties that would contribute to this measurement.

\paragraph*{Prospects for Theoretical Uncertainties:}

In this section we highlight several theoretical issues related to achieving an accurate description of the $N_2$ and jet mass spectra. The resolution of these issues is well beyond the scope of the current paper, and our goal is therefore more to emphasize where progress can be made, and what issues must be overcome.  From \Fig{fig:dependence}, we see that perturbatively there is an extremely weak dependence of the jet mass on the $N_2$ cut, and that the jet mass aligns well with the $W$ or $Z$ world average. The small negative offset is due to radiation that is not captured in the jet, due to the finite jet radius. These effects can be analytically calculated, and are incorporated in all standard jet substructure calculations. We therefore believe that perturbative effects can be kept under good theoretical control.

More concerning are non-perturbative effects, which, as can be seen in \Fig{fig:dependence} dominate the offset of the jet mass from the $W$ and $Z$ world average, and furthermore, exhibit a dependence on the $N_2$ cut. In fact, there are two distinct non-perturbative effects which would need to be understood in order to completely understand this measurement. The first are non-perturbative corrections to the $N_2$ distribution on which the cut is applied.  Non-perturbative effects for the groomed $D_2$ observable \cite{Larkoski:2014gra,Larkoski:2015kga} (which is closely related to $N_2$) were recently studied in \cite{Larkoski:2017iuy,Larkoski:2017cqq,Moult:2017okx}, where it was shown that they take a relatively simple form, and can be modeled by a single parameter shape function. The second is non-perturbative corrections to the jet mass distribution itself. Non-perturbative corrections to the groomed top quark mass distribution were studied in \cite{Hoang:2017kmk}, where they were also found to take a simple form. However, non-perturbative effects for the mass distribution for the decay of a color singlet have not, to our knowledge, been studied in the literature. We believe that this deserves further attention. Ideally, these corrections could also be described by a universal shape function that could then be self-consistently extracted with the mass measurement itself.

As a cause for cautious optimism, we would like to point out that for the $m_W - m_Z$ measurement strategy, it is not the non-perturbative corrections themselves that will need to be understood at the 10\% level, but rather their difference acting on $Z$ and $W$ bosons given particular selection criteria. Thus, the $\mathcal{O}(100 \text{ MeV})$ effects quoted should not be interpreted as requiring control over absolute hadronization and underlying event corrections at the single hadron level, which is unrealistic. Ideally, one could therefore prove a statement on the universality of the non-perturbative corrections for hadronic $W$ and $Z$ decays, which would place this measurement on a firmer theoretical footing. This universality would be violated by, for example, $b$-quark mass effects, but this should be a much smaller effect than the overall shift due to hadronization, and could perhaps be accounted for. Therefore, while this measurement seems challenging from a theoretical perspective, it points to a number of theoretical issues which deserve further thought, and whose resolution would have wider applicability in a number of jet substructure measurements.

\section{Conclusions and Outlook}
\label{sec:summary}

A feasibility study for a first measurement of the $W$ boson mass in the all-jets final states at the LHC and HL-LHC has been presented.
Compared to the lepton plus neutrino final state, a measurement in the all-jets final state could avoid experimental systematic uncertainties related to the measurement of the missing transverse momentum and the theoretical uncertainties related to the transverse mass.
While a measurement of the $W$ mass itself seems unrealistic since it would require a significantly better understanding of the jet energy calibration than reached by the current LHC experiments, a measurement of the mass difference between the $W$ and $Z$ bosons is more feasible.
New trigger strategies will need to be exploited to reach a statistical uncertainty of 30 MeV with HL-LHC data corresponding to 3000/fb of integrated luminosity.
The measurement is, however, limited by the understanding of non-perturbative contributions to the invariant masses of $W\to q\bar{q}'$ and $Z\to b\bar{b}$, that would need a significant improvement to reach below 100 MeV precision.

More generally, we believe that progress towards the extraction of the $W$ mass from the all hadronic final state using jet substructure represents a concrete goal that can drive progress in jet substructure, much like the extraction of $\alpha_s$ from the jet mass distribution in hadron colliders \cite{Bendavid:2018nar}. We have highlighted areas for improvement on both the theoretical and experimental sides. Their improved understanding will have a much broader impact on jet substructure, most importantly for improving searches for light hadronically decaying resonances, which utilize many of the same techniques, and almost certainly for other applications unforeseen at the current time.

\begin{acknowledgments}

We thank Andrew Larkoski for useful discussions and collaboration at early stages of this project. We thank Ben Nachman for useful discussions.
MF is supported by the U.S.\ Department of Energy (DOE) under contract DE-SC0011640.
AH gratefully acknowledges funding in the Emmy-Noether program (HI 1952/1-1) of the German Research Foundation DFG. PH acknowledgese the support of the MIT Physics department.
IM is supported by the DOE under grant number DE-AC02-05CH11231 and the LBNL LDRD program.
NT and CV are supported by the Fermi Research Alliance, LLC under Contract No. DE-AC02-07CH11359 with the DOE, Office of Science, Office of High Energy Physics. 

\end{acknowledgments}

\clearpage
\appendix
\section{$W$ and $Z$ Jet Mass Distributions in Pythia and Herwig}
\label{sec:appendix}

\begin{figure}[htb]
\begin{center}
\includegraphics[width=0.45\linewidth]{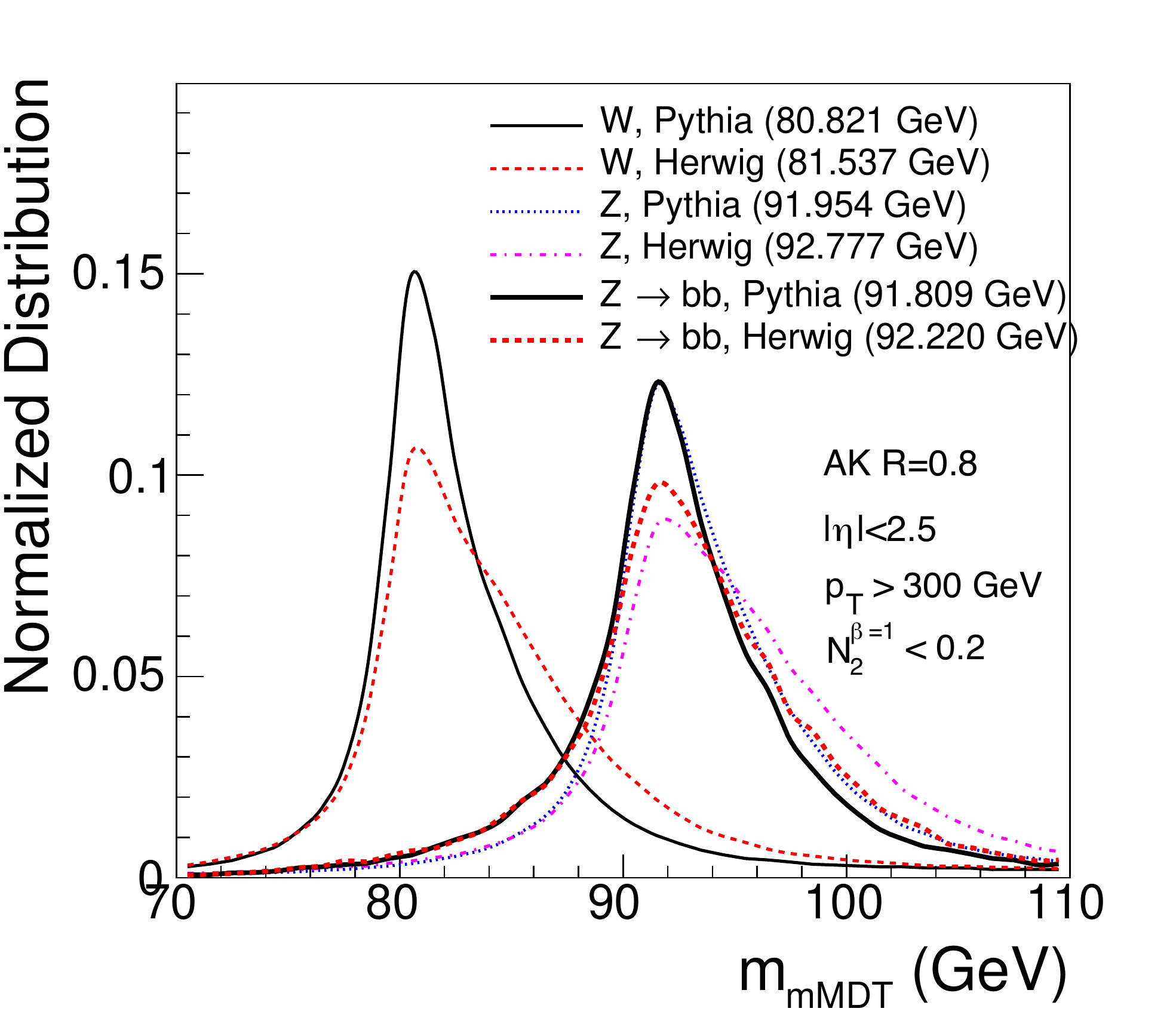}
\includegraphics[width=0.45\linewidth]{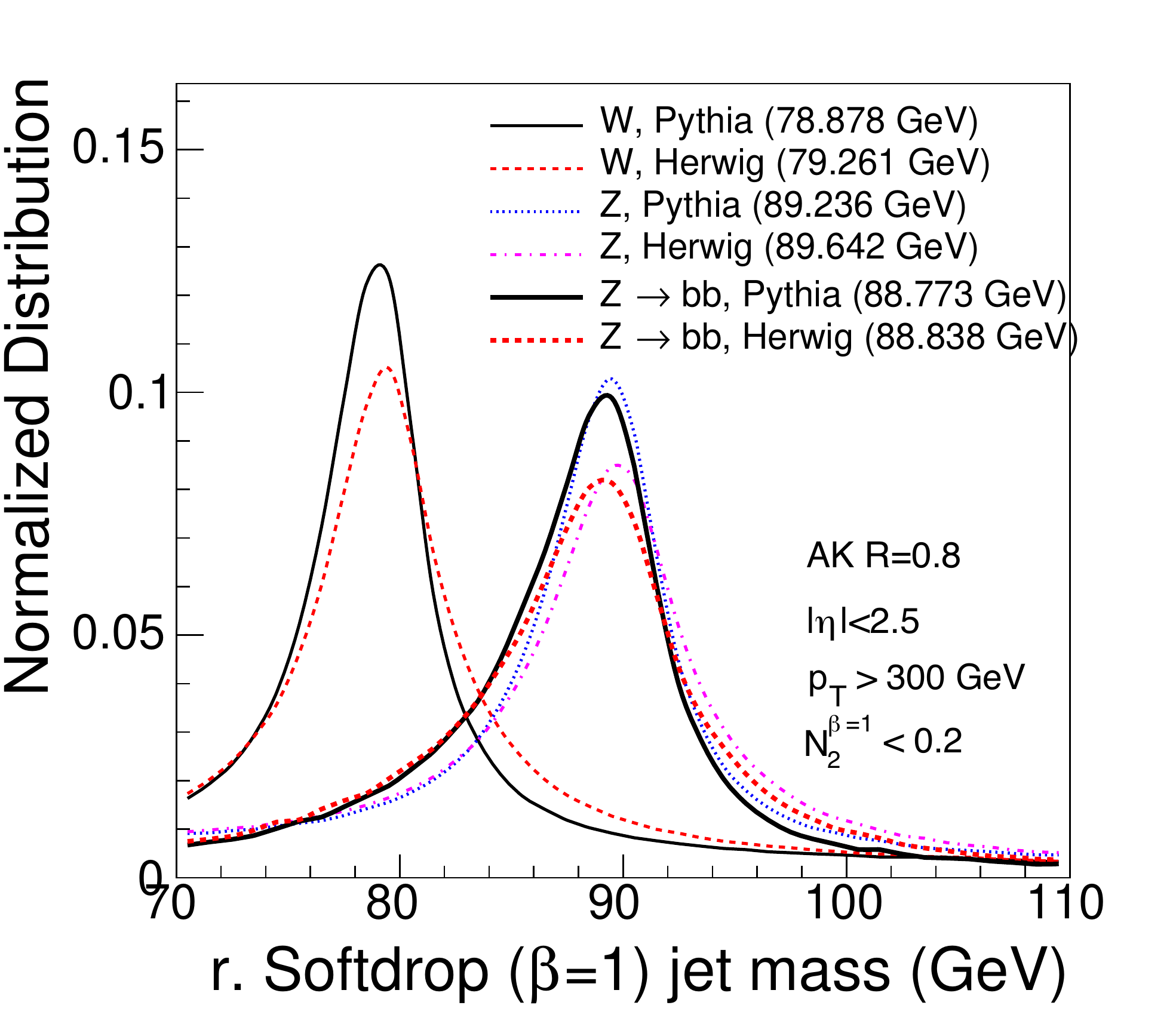} \\
\includegraphics[width=0.45\linewidth]{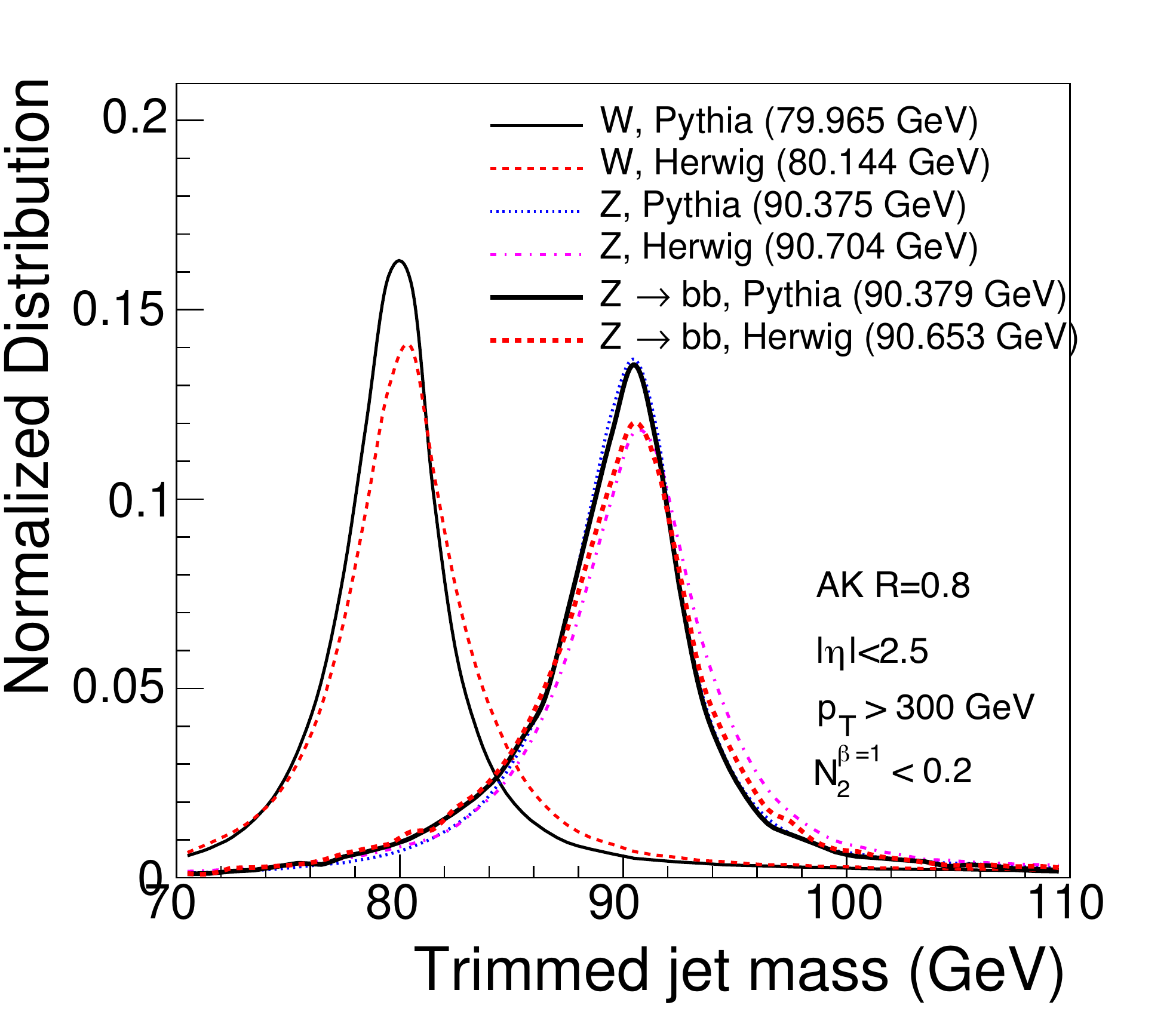}
\includegraphics[width=0.45\linewidth]{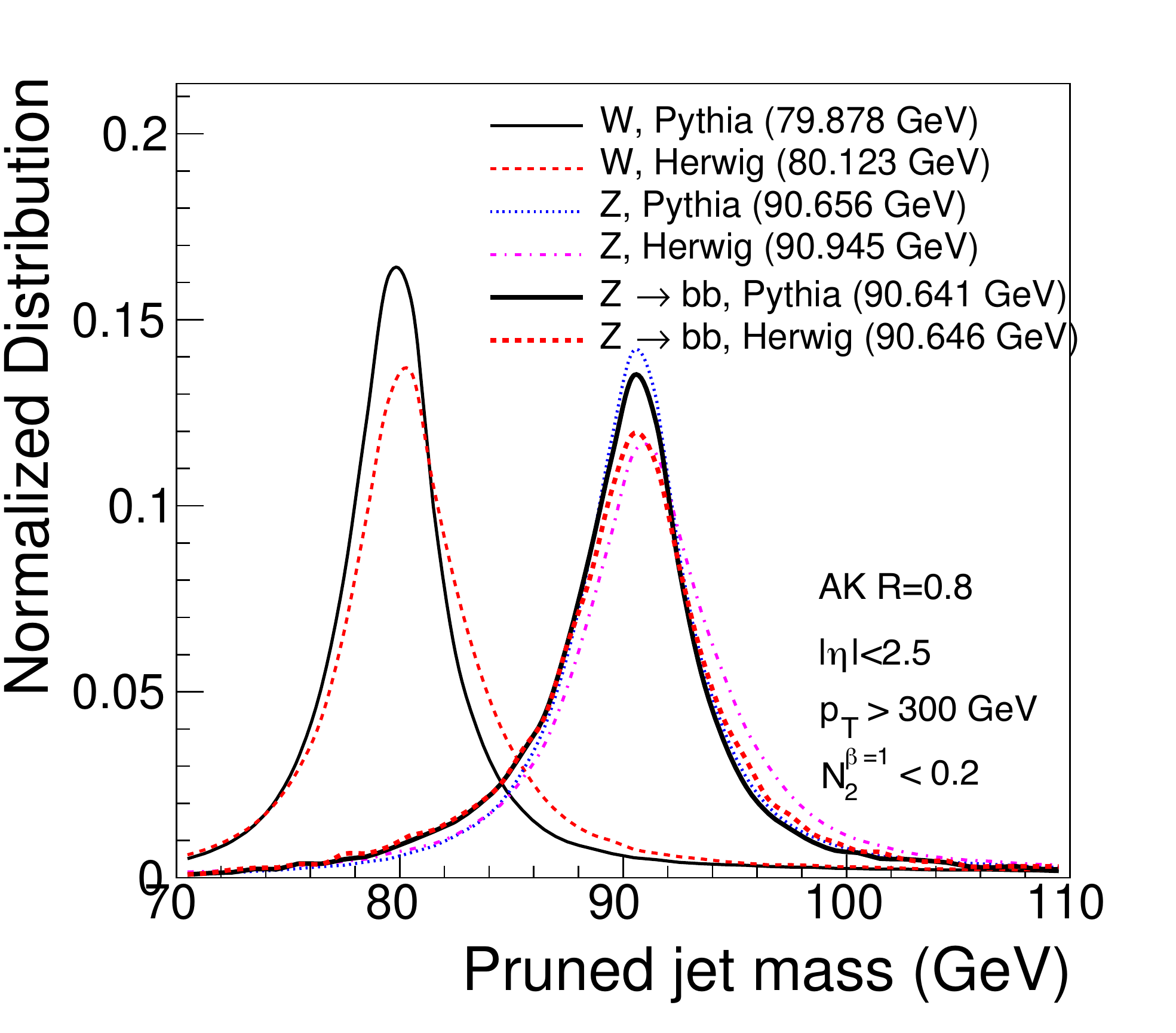}
\end{center}
\caption{Jet mass distribution of $W$, $Z$ and $Z\to b\bar{b}$ jets in \textsc{pythia} and \textsc{herwig} with \pt larger than 300 GeV and N$_2^{\beta=1}<0.2$ for four different grooming algorithms.
(top left) mMDT with the angular exponent $\beta = 0$, soft cutoff threshold $z_\mathrm{cut} = 0.1$, and characteristic radius $R_0 = 0.8$.
(top right) recursive softdrop~\cite{Dreyer:2018tjj} with the angular exponent $\beta = 1$, soft cutoff threshold $z_\mathrm{cut} = 0.1$, characteristic radius $R_0 = 0.8$, and the number of iterations N set to infinity.
(bottom left) trimming~\cite{Krohn:2009th} with subjet size of $R_{\text{sub}}=0.2$ and $f_{\text{cut}=0.03}$.
(bottom right) pruning~\cite{Ellis:2009me} with the soft threshold parameter $z_{\text{cut}}=0.1$ and angular separation threshold of $\Delta R > m_{\text{jet}}/p_{\text{T,jet}}$. }
\label{fig:dependence2}
\end{figure}

\begin{figure}[htb]
\begin{center}
\includegraphics[width=0.45\linewidth]{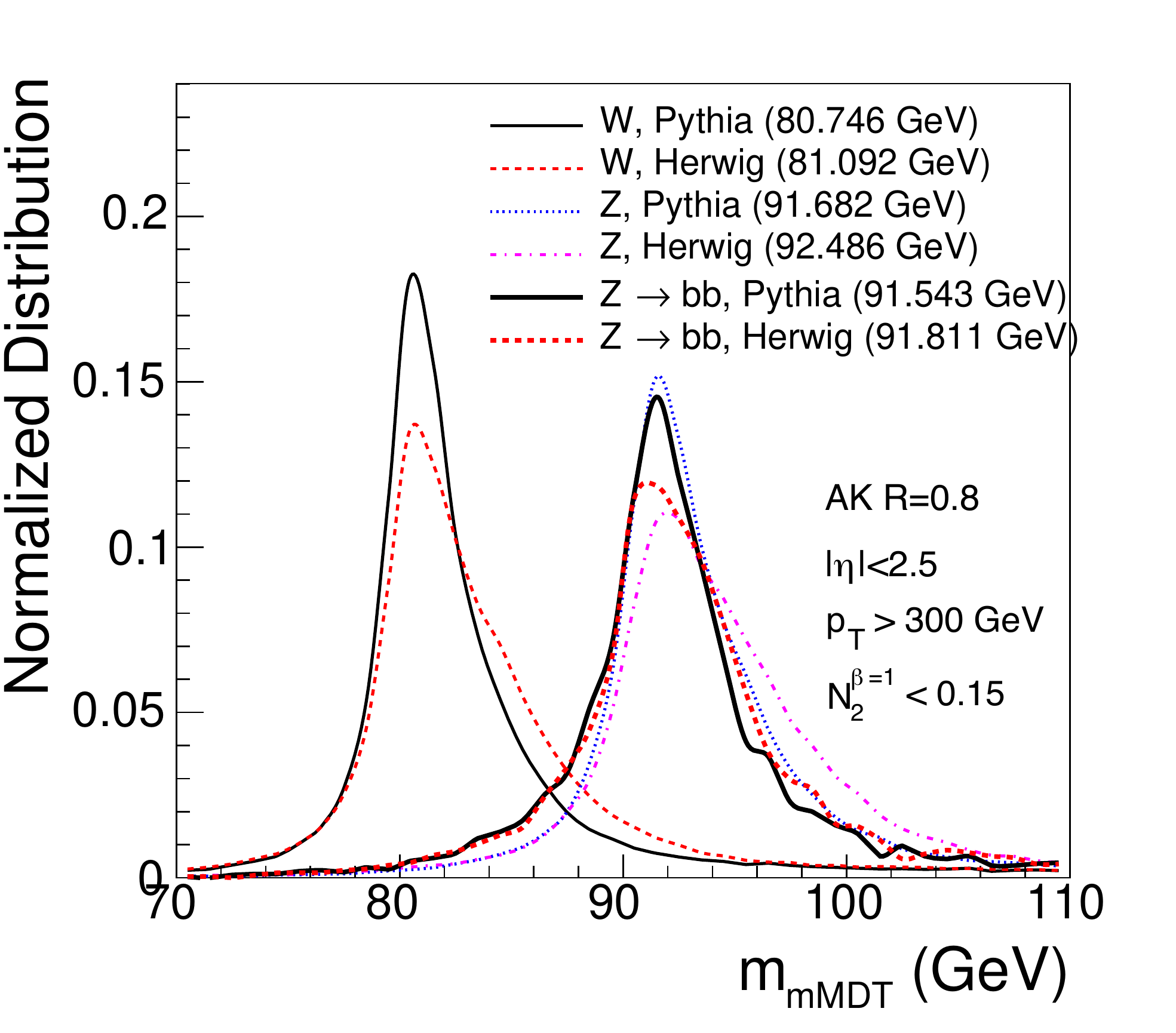}
\includegraphics[width=0.45\linewidth]{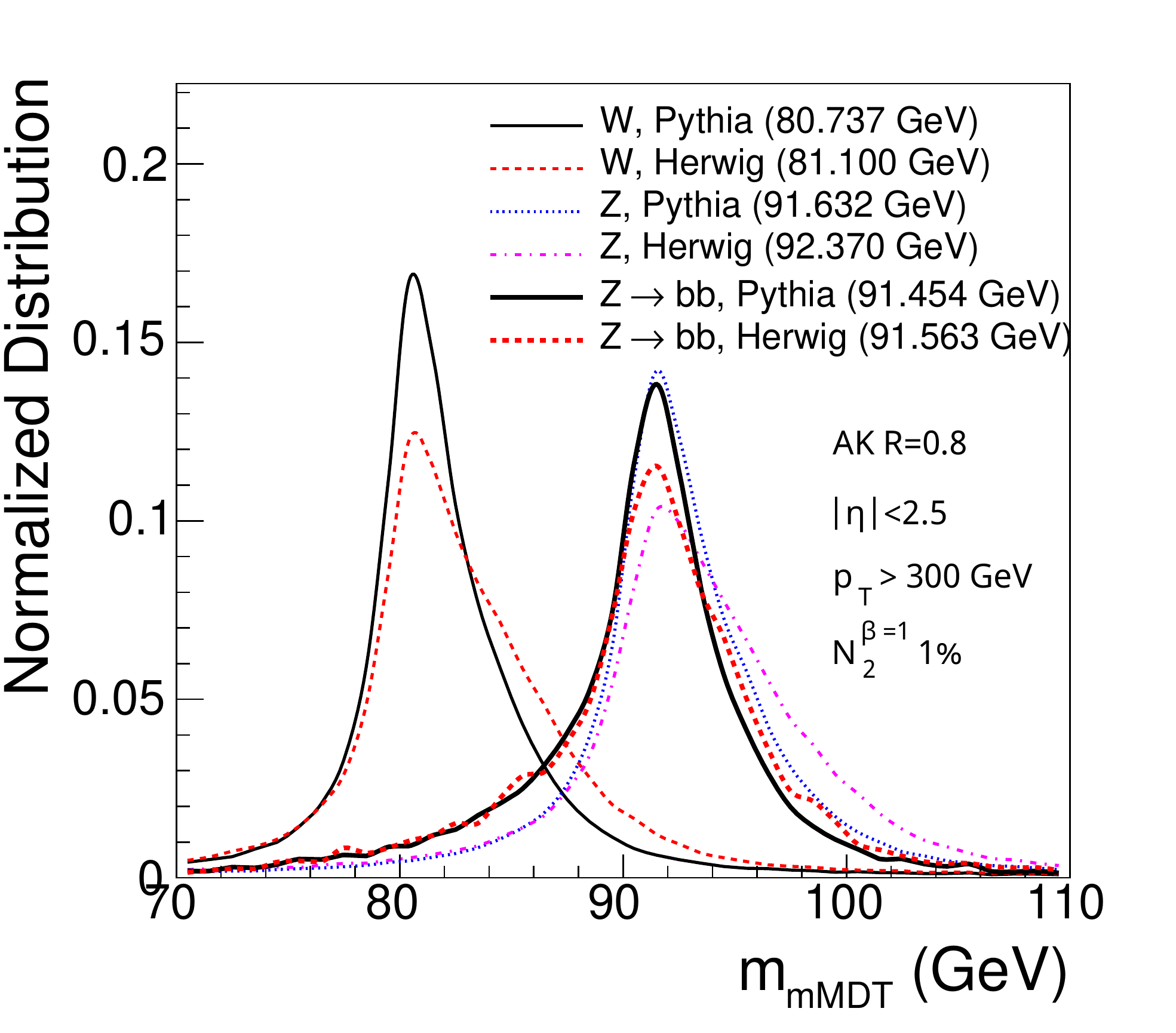} \\
\includegraphics[width=0.45\linewidth]{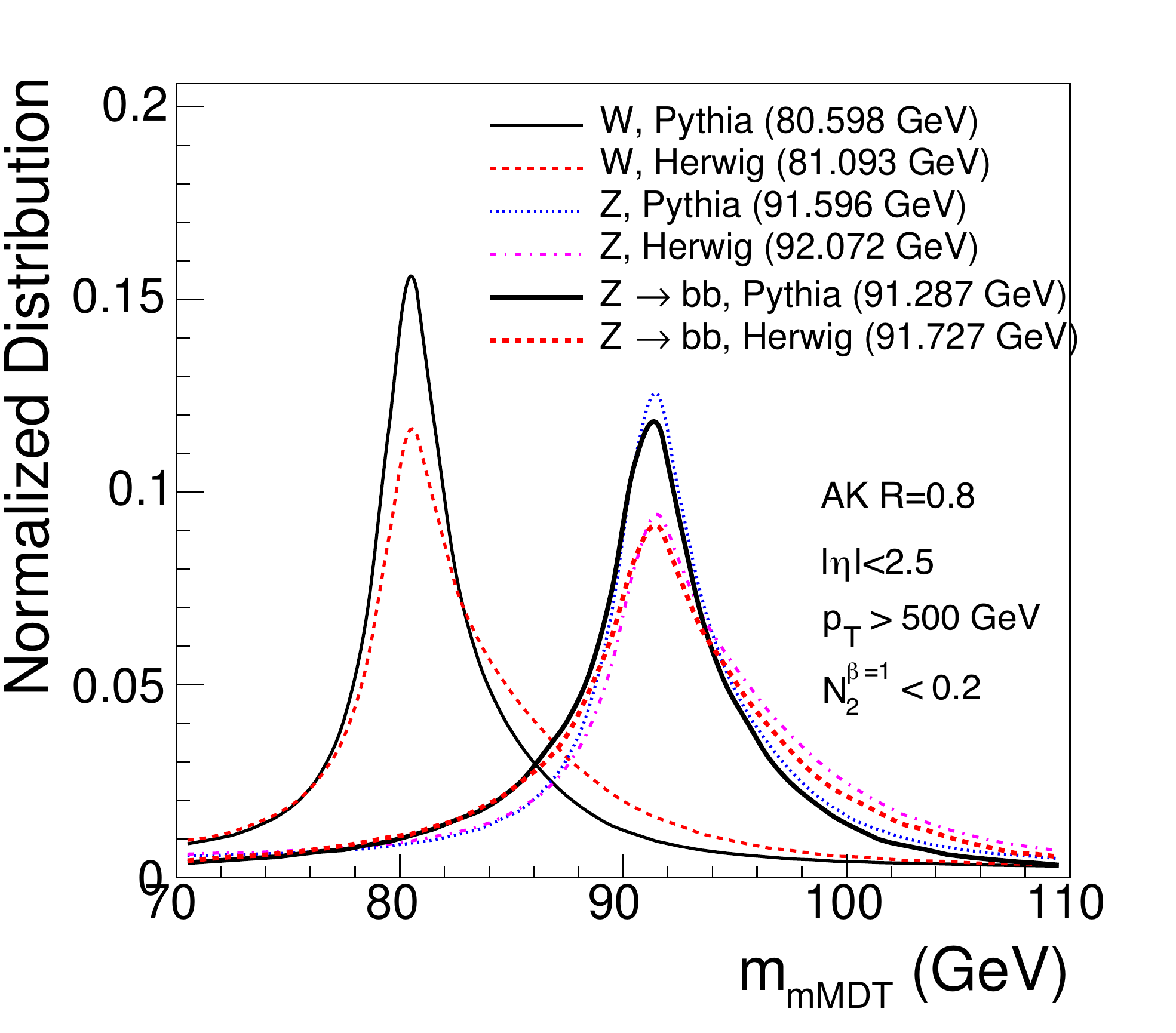}
\end{center}
\caption{Jet mass distribution of $W$, $Z$ and $Z\to b\bar{b}$ jets in \textsc{pythia} and \textsc{herwig} using mMDT with the angular exponent $\beta = 0$, soft cutoff threshold $z_\mathrm{cut} = 0.1$, and characteristic radius $R_0 = 0.8$ for different selections on \pt and N$_2^{\beta=1}$.
(top left)  N$_2^{\beta=1}<0.15$ and \pt larger than 300 GeV.
(top right) decorrelated $N_2^{\beta=1}$ 1\% and \pt larger than 300 GeV.
(bottom) N$_2^{\beta=1}<0.2$ and \pt larger than 500 GeV.}
\label{fig:dependence3}
\end{figure}

In Fig.~\ref{fig:dependence2} and Fig.~\ref{fig:dependence3} we provide the predictions from \textsc{pythia} and \textsc{herwig} for different grooming algorithms, transverse momentum thresholds and substructure observable selections.
They demonstrate how the dependence of the shape of the jet mass observable on non-perturbative effects, parton shower and hadronization can be influenced by the choice of grooming algorithm, transverse momentum threshold and substructure observable selection.
It can be seen that the peak positions, symmetry of the distributions and the differences between \textsc{pythia} and \textsc{herwig} depend evidently on the choice grooming algorithm.

\newpage
\bibliographystyle{JHEP}
\bibliography{mWhadronic}

\end{document}